\documentclass[journal]{IEEEtran}

\usepackage{cite}
\usepackage{forest}
\usepackage{amsmath}
\usepackage{algorithmic}
\usepackage{cite}
\usepackage{amsmath,amssymb,amsfonts}
\usepackage{pifont}
\newcommand{\xmark}{\ding{55}}%
\usepackage{algorithmic}
\usepackage{graphicx}
\usepackage{textcomp}
\usepackage{xcolor} 
\usepackage{siunitx}
\usepackage{booktabs}
\usepackage{tabularx}
\usepackage{ctable}
\usepackage{multirow}
\usepackage{float}
\usepackage{caption}
\usepackage{color}
\usepackage[normalem]{ulem}
\usepackage{tabularray}
 \usepackage{stfloats}

\begin{document}	
	
	\title { A Survey on Security of Ultra/Hyper Reliable Low Latency Communication: Recent Advancements, Challenges, and Future Directions}

\author{Annapurna Pradhan,~\IEEEmembership{Member,~IEEE},
		Susmita Das,~\IEEEmembership{Senior Member,~IEEE}, Md. Jalil Piran,~\IEEEmembership{Senior Member,~IEEE}, and Zhu Han,~\IEEEmembership{Fellow,~IEEE} 

 \thanks{A. Pradhan and S. Das are with the Department of Electrical 
         Engineering, National Institute of Technology, Rourkela, India, (email: pradhan.annapurna37@gmail.com and sdas@nitrkl.ac.in).}
\thanks{M. J. Piran is with the Department of Computer Science and 
         Engineering, Sejong University, Seoul 05006, South Korea, (email: piran@sejong.ac.kr).}
\thanks{Z. Han is with the Department of Computer Science, University of Houston, Houston, TX 77004, USA (email: zhan2@uh.edu).}         
}

\maketitle
	

\begin{abstract}
Ultra-reliable low latency communication (URLLC) is an innovative service offered by fifth-generation (5G) wireless systems. URLLC enables various mission-critical applications by facilitating reliable and low-latency signal transmission to support extreme Quality of Service (QoS) requirements. Apart from reliability and latency, ensuring secure data transmission for  URLLC has been a prominent issue for researchers in recent years. Using finite blocklength signals to achieve the stringent reliability and latency criteria in URLLC eliminates the possibility of using conventional complex cryptographic security enhancement techniques based on encoding and decoding of secret keys. Thus, the development of lightweight security mechanisms is of paramount importance for URLLC. Recently, Physical-Layer Security (PLS) techniques have emerged as a powerful alternative to the complex cryptography-based security approaches for facilitating secure URLLC by exploiting the randomness of the wireless channel. Therefore, in this survey, we present a comprehensive and in-depth review of the state-of-the-art PLS enhancements utilized to unleash secure URLLC while analyzing the impact of various system design parameters on its performance. Moreover, the survey incorporates a detailed overview of the recent advancements in ensuring secure URLLC using PLS in various mission-critical applications, and 5G URLLC enabling technologies like non-orthogonal multiple access (NOMA), multi-antenna systems, cooperative communication using unmanned aerial vehicles (UAV), and intelligent reflective surfaces (IRS). Apart from this, we briefly discuss the role of advanced Machine Learning (ML) techniques in designing robust and intelligent PLS schemes for URLLC service. Then, for the first time, the survey introduces the extended service class of URLLC in 6G, i.e., Hyper Reliable Low Latency Communication (HRLLC), and provides an outlook on the future security aspects while identifying some promising new technologies that can provide secure HRLLC in 6G. We also identify several key challenges and open issues faced by URLLC in achieving the desired security levels from the physical layer perspective. Finally, the survey highlights interesting future research directions for designing robust, efficient, and optimal PLS techniques to support the URLLC service in 5G and HRLLC in future 6G wireless networks.

\end{abstract}
	
	\begin{IEEEkeywords}
		6G, Hyper Reliable Low Latency Communication (HRLLC), URLLC, Physical Layer Security, NOMA, UAV, IRS, Machine Learning.
	\end{IEEEkeywords}

	\IEEEpeerreviewmaketitle
		
	\section{Introduction}
		
	\IEEEPARstart{T}{he} introduction of the fifth-generation (5G) wireless systems has been a game changer in enabling many futuristic applications and services. The ambitious vision of 5G of connecting the world through autonomous networks has set the foundation for the future sixth-generation (6G) wireless systems \cite{pls72}. This brings 5G to introduce novel services such as enhanced Mobile Broadband (eMBB), Ultra-Reliable Low Latency Communication (URLLC), and massive Machine Type Communication (mMTC) to deal with the growing demand for smart applications \cite{pls99}. Out of all, URLLC is the promising new service offered by 5G that supports mission-critical applications like autonomous driving, industrial automation, and smart healthcare systems by enabling high reliability (99.999$\%$) and low latency ($\leq 1 ms$) signal transmission \cite{pls78}, \cite{pls95}. For this, URLLC data transmission uses short packet finite blocklength signals to satisfy the stringent reliability and latency constraints simultaneously. Apart from this, URLLC brings significant novelty to 5G by introducing the qualitative difference from the previous generation of mobile services while expanding functionality and transcending the boundaries of new-age 5G applications.

    URLLC provides seamless connectivity to a massive number of Internet of Things (IoT) devices for enabling smart applications through transformative technologies that rely on real-time and highly dependable communication \cite{r101}. Generally, URLLC service is responsible for transmitting confidential and high-priority control information in various mission-critical 5G use cases like autonomous driving and industrial automation scenarios \cite{r56}. Moreover, the exponential increase in service demand and the open nature of the wireless medium make URLLC signals vulnerable to security threats which leads to information leakage \cite{r4}. Therefore, ensuring secure communication becomes more challenging as URLLC service scales up to accommodate this exponentially growing service demand. Thus, ensuring secure communication becomes a crucial requirement to protect these sensitive URLLC data and prevent information leakage. 

Traditionally, the design of security enhancement schemes relies on an information-theoretic approach where cryptographic techniques were adopted to prevent information leakage and establish secure communication in wireless systems \cite{r104}, \cite{r14}. However, these schemes are basically suitable for signals with larger blocklength. On the other hand, for URLLC, the signal blocklength is small and finite to satisfy the low-latency criteria. Again, the stringent reliability and latency constraints of URLLC put an additional burden on the design of security enhancement schemes. Moreover, the finite blocklength introduces a non-zero decoding error probability (DEP), which complicates the design of security enhancement schemes for URLLC \cite{pls79}.  Therefore, the complex cryptographic or key-based secrecy techniques cannot be used for URLLC as they consume extra processing time and incur signaling overheads \cite{r116}. 

Meanwhile, to address these challenges, physical layer security (PLS) has emerged as a potential technique for providing security to URLLC signal transmission.  PLS exploits the randomness of the wireless channel characteristics to ensure secure communication, making it difficult for eavesdroppers to intercept the transmitted data \cite{r5}.In fact, PLS provides flexible and low-complex security enhancements when compared to complex cryptographic methods. It ensures secure message transmission without the need to generate additional secret keys and complex encryption/decryption techniques. Compared to higher-layer cryptographic solutions, PLS generates less signaling overheads by reducing additional bandwidth requirements while utilizing less computational resources and processing power. This is particularly beneficial for IoT applications containing resource-constrained devices by conserving system resources \cite{r42}.  

	Additionally, conventional cryptographic security techniques may be vulnerable to attacks that can exploit the designed algorithms in the key-based security management systems. However, PLS techniques are more robust against jamming and eavesdropping than conventional cryptographic schemes \cite{r50}. This ensures secure communication even in the presence of intentional disruptions. Apart from this, PLS can adapt to dynamic wireless channel conditions, making it suitable for time-varying environments like high mobility scenarios in vehicular communication. 
 
 PLS in URLLC offers unique advantages in terms of security, improved reliability, reduced latency, and resource efficiency compared to conventional cryptographic techniques.  The state-of-the-art recent studies unveil the potential benefits of using PLS techniques for providing secure URLLC in 5G. It provides secure communication in URLLC mission-critical applications without compromising the QoS requirements. From these studies, it is well evidenced that, PLS can significantly enhance the security of URLLC signal transmission while protecting against potential threats to ensure data confidentiality and security.

   \begin{figure}[t]
	\centerline {\includegraphics[width=\linewidth, height=8.8 cm]{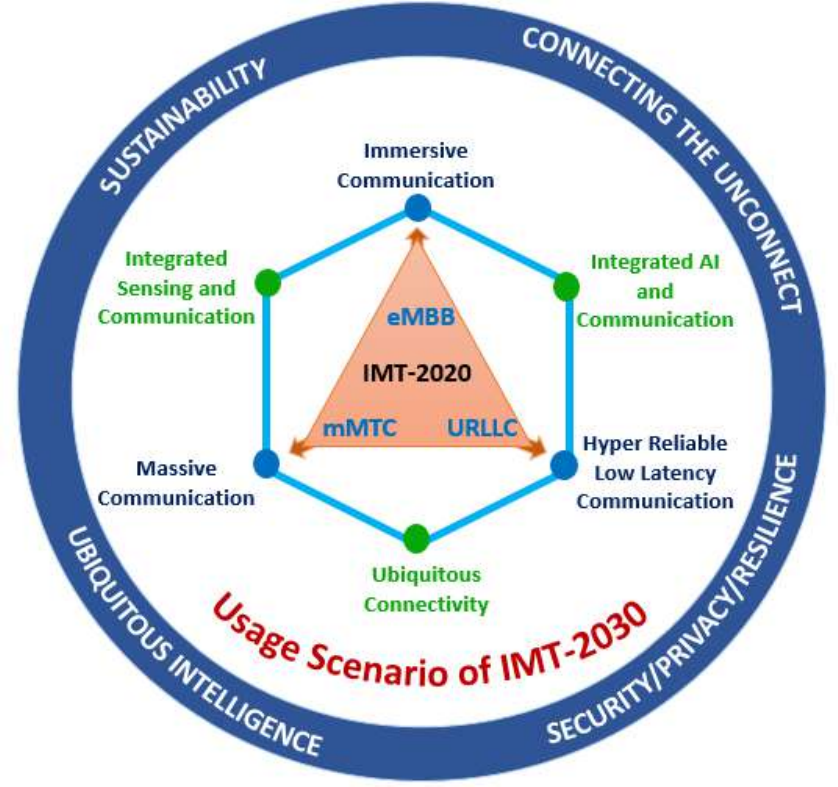}}
	\caption{An illustration of IMT-2030 recommended Usage Scenarios}
	\label{fig12}
	\vspace{-4 mm}
\end{figure}

   Due to the ever-growing service demand, dynamic network topology, and revolutionized smart 5G applications, ensuring URLLC becomes difficult and is expected to face severe concerns regarding unknown threats. Hence, to provide seamless secure URLLC service, flexible and intelligent security techniques are needed which can protect against eavesdropping while capturing the wireless channel dynamics. To this end, the amalgamation of machine learning (ML) techniques and PLS can bring groundbreaking steps toward defining an adaptive and intelligent security framework while optimizing resource allocation and spectrum usage when extending it to provide secure URLLC at the network edge. Incorporating ML-driven PLS framework into URLLC applications will provide robust security while establishing and achieving multi-layer intelligence for URLLC with proactive mitigation of security threats \cite{pls86}.    Recent research on ML-enabled URLLC has demonstrated the efficacy of using ML models in optimizing the network function and guaranteeing the end-to-end QoS through intelligent scheduling and resource allocation, optimizing beamforming and congestion control. Although ML brings potential benefits in achieving URLLC QoS in various application scenarios, security threats at the intermediate gradients of the ML models can increase the risk of information leakage and disclosure of sensitive information. Therefore, developing secure and privacy-preserving ML models may pave the way to achieving intelligent and secure URLLC.

    \begin{table}[t!]
    \caption{List of Acronyms}
    	\label{tab1}
    	\begin{tabular}{|l|l|}
    		\hline
    		
    		\textbf{ Acronym  }   & \textbf{ \,\,\,\,  Definition }  \\ \hline

           5G           & Fifth Generation     \\  \hline
			6G           & Sixth Generation     \\ \hline
			3GPP         & Third Generation Partnership Project  \\  \hline
			AI           & Artificial Intelligence      \\ \hline	
			AP           & Access Point        \\  \hline
			AN           & Artificial Noise   \\  \hline
			AR           & Augmented Reality   \\  \hline
			BS           & Base Station        \\  \hline	
			CNN          & Convolutional Neural Network  \\  \hline	
			CSI          & Channel State Information    \\   \hline
			COP          & Connection Outage Probability   \\ \hline
			D2D          & Device-to-Device	  \\ \hline
			DEP          & Decoding Error Probability 	  \\  \hline
			DL           & Deep Learning	  \\  \hline
			DNN          & Deep Neural Network   \\  \hline
			DRL          & Deep Reinforcement Learning   \\  \hline
			eMBB         & Enhanced Mobile Broadband 	\\  \hline
			EE           & Energy Efficiency	\\  \hline
			FBL          & Finite Block Length	\\  \hline
			FL           & Federated Learning	\\  \hline
			IoT          & Internet of Things	\\  \hline
			IIoT         & Industrial Internet of Things	\\  \hline
			IRS          & Intelligent Reflective surfaces	\\  \hline
			ITU          & International Telecom Union  \\  \hline
			LoS          & Line-of-Sight	\\  \hline    		
			LTE          & Long Term Evolution	 \\  \hline    		
			MAC          & Medium Access Control \\  \hline
			MEC          & Mobile Edge Computing \\  \hline
			MIMO         & Multiple Input Multiple Output \\  \hline
			mMIMO         & Massive Multiple Input Multiple Output \\  \hline
			ML           & Machine Learning     \\  \hline
			mMTC         & Massive Machine Type Communication  \\ \hline
			NN           & Neural Network            \\  \hline
			NOMA         & Non-Orthogonal Multiple Access \\ \hline
			NLoS         & Non-Line-of-Sight      \\  \hline
			OMA          & Orthogonal Multiple Access\\  \hline
			PLS          & Physical Layer Security \\   \hline    		
			QC           & Quantum Computing         \\   \hline
			QoS          & Quality of Service       \\   \hline    			
			RL           & Reinforcement Learning    \\   \hline    	
			RRM          & Radio Resource Management     \\   \hline
			SG	         & Secrecy Gap      \\   \hline
			SL           & Supervised Learning      \\   \hline
			SNR          & Signal-to-Noise Ratio     \\   \hline
			SOP          & Secrecy Outage Probability \\   \hline
			SPC          & Short Packet Communication\\   \hline    		
			UE           & User Equipment            \\   \hline
			UAV          & Unmanned Ariel Vehicle    \\   \hline
			UL           & Un-supervised Learning       \\  \hline
			VR           & Virtual Reality       \\  \hline
			WSN          & Wireless Sensor Network   \\   \hline
			XR           & Extended Reality       \\  \hline

    	\end{tabular}
    	\centering
    	
    	\vspace{-4 mm}
    \end{table}   

  With the commercial deployment of the 5G system, wireless communication has transformed in an unprecedented manner over the past decade. This remarkable growth in connectivity support and integration of new capabilities for communication and sensing escalates the ongoing evolution of the communication landscape towards 6G. Recently, ITU has released its recommendations for implementing 6G under the vision 2030, called International Mobile Telecommunications (IMT)-2030 \cite{recommendation2023framework}. The recommended IMT-2030 is an extension of the current IMT-2020 capabilities, which are curated based on the transformational changes experienced so far. As depicted in Fig. \ref{fig12}, the 5G services will be extended in IMT-2030 recommendations that define the new usage scenarios where URLLC will be enhanced to Hyper Reliable Low Latency communication (HRLLC), eMBB will be termed as immersive communication with advanced capabilities, and mMTC will be transformed into massive communication in 6G \cite{recommendation2023framework}. Apart from this, three more usage scenarios such as integrated sensing and communication, integrated AI and communication, and ubiquitous connectivity are introduced based on the the envisioned proliferation of new applications and service demand in 6G.  

   6G is expected to encompass a wide range of mission-critical applications with diverse QoS requirements including IoT-based industrial automation to critical data transmission in smart healthcare and autonomous driving applications. The unprecedented revolution of emerging applications like tactile internet, holographic telepresence, brain-computer interaction, and extended reality (XR) is also expected to be supported by HRLLC in 6G \cite{pls93}.  Here, the large-scale network integration in these applications engages in confidential information exchange and builds optimized networks using artificial intelligence (AI) to manage and support the large volume of generated data. However, the future 6G HRLLC applications will substantially impose strict reliability, latency, and security requirements to satisfy the diverse QoS requirements of heterogeneous users. In such cases, only utilizing the dynamic characteristics of wireless channels to ensure security at the physical layer will not be enough. Moreover, setting up a PLS scheme for HRLLC operations will be even more difficult because of evolving hardware design, communication protocols, the large volume of data traffic, and extreme QoS constraints in 6G. Therefore, designing flexible PLS schemes, integrating AI-enhanced security, and using advanced techniques like quantum communication and blockchains can proactively detect and mitigate potential threats to fortify the uninterrupted and secure HRLLC against cyber-attacks in 6G.

\subsection{Related Surveys}

\subsubsection{Survey and Tutorials on PLS}

 PLS improvement strategies and enabling schemes in 5G have been comprehensively discussed in recent years. We thoroughly investigated the related survey works on PLS presented in \cite{pls1}, \cite{pls71}, \cite{pls21}, \cite{pls23} \cite{r105}, \cite{r109}, \cite{r110} and discussed various aspects of security issues, potential threat models, enabling technologies, and research gaps in providing secure wireless communication. Specifically, the survey in \cite{r105} presented a comprehensive survey on PLS techniques used for 5G IoT networks. The work also identified and discussed various security threats affecting physical layer communication and corresponding countermeasures adopted in 5G IoT application scenarios. Another survey work in \cite{r109} discussed the PLS enhancement schemes adopted in various 5G technologies like multiple-input and multiple-output (MIMO) antenna systems, non-orthogonal multiple access (NOMA), millimeter wave (mmWave) communication, and heterogeneous networks while identifying the inherent challenges in ensuring secure communication in each of the scenarios. Similar to the aforementioned surveys, the work in \cite{r110} also presented some of the key concepts of PLS techniques, security threats, and countermeasures for 5G wireless networks and the recent advances in various 5G technologies for evaluating the PLS performance for long blocklength codes. 
 

\subsubsection{Survey and Tutorials on URLLC}

Recently, we have witnessed an increase in the number of research work related to the key enabling technologies for URLLC service in 5G and beyond networks. For example, the surveys in \cite{r8}, \cite{r153}, and \cite{r154} have extensively discussed various enabling technologies, interference management schemes, and resource management challenges under stringent QoS requirements for URLLC service. Specifically, the work in \cite{r153} discussed various interference management issues for URLLC service in a heterogeneous wireless network. In \cite{r154}, the authors have briefly discussed the scheduling and resource allocation techniques adopted in recent research works for URLLC service. Similarly, one of the surveys in \cite{r159} has provided a comprehensive study of the various URLLC enabling techniques from the physical and MAC layer perspective. The work also identifies some of the specific emerging research areas to be focused on for developing efficient transmission schemes considering reliability and latency constraints of finite blocklength URLLC signal. Moreover, in \cite{r77}, a detailed overview of the URLLC, its features, and QoS requirements are presented as given in 3GPP 5G new radio (NR) Release 15 and Release 16. Additionally, an outlook from 3GPP Release 17 is also discussed for the utilization of unlicensed spectrum targeting further enhanced URLLC operation. 

Most importantly, none of the above surveys discuss the security issues for finite blocklength URLLC signal transmission. Moreover, the aforementioned surveys lack an in-depth analysis of the physical layer parameters like signal blocklength, pilot signal length, CSI estimation error, and dense network scenarios that can potentially affect overall PLS performance. However, these survey works fail to discuss the security challenges and their impact on URLLC signal transmission in mission-critical applications.


\begin{table*}[t!]
\caption{Comparison of the State-of-the-art Surveys on PLS and URLLC in wireless communication}
\label{tab2}
\begin{center}
\begin{tabular}{|c|c|l|c|c|c|c|c|c|}
\hline
Research  & Year & Survey Objective & \begin{tabular}[c]{@{}c@{}}PLS of \\ Infinite \\ blocklength\end{tabular} & \begin{tabular}[c]{@{}c@{}}PLS of \\ Finite \\ blocklength\end{tabular} & \begin{tabular}[c]{@{}c@{}}PLS \\ Enablers\end{tabular} & \begin{tabular}[c]{@{}c@{}}ML-based\\ Security\end{tabular} & \begin{tabular}[c]{@{}c@{}}PLS in \\ 6G\end{tabular} & \begin{tabular}[c]{@{}c@{}} Challenges and \\ Future Scope \end{tabular}\\ \hline

\cite{pls22}   & 2015 & \begin{tabular}[c]{@{}c@{}}PLS in Wireless Communication\end{tabular} & \checkmark    & \xmark      & \checkmark    & \xmark     & \xmark     & \checkmark     \\ [1mm] \hline

\cite{pls21}   & 2017 & PLS of 5G networks  & \checkmark   & \xmark     & \checkmark    & \xmark    & \xmark   & \checkmark   \\ [1mm] \hline

\cite{r109}  & 2018 & PLS of 5G networks  &\checkmark     & \xmark   & \checkmark   & \xmark     & \xmark   & \checkmark    \\ [1mm] \hline

\cite{r7}   & 2019 & \begin{tabular}[c]{@{}c@{}}Optimization Approach for PLS\end{tabular} & \checkmark   & \xmark  & \checkmark       & \xmark   & \xmark   & \checkmark   \\  [1mm] \hline

\cite{pls1}  & 2019 & PLS of 5G networks  & \checkmark     & \xmark   & \checkmark   & \xmark     & \xmark    & \checkmark   \\  [1mm] \hline

\cite{r105}   & 2019 & \begin{tabular}[c]{@{}c@{}}PLS of 5G IoT\end{tabular}     & \checkmark   & \xmark   & \checkmark  & \xmark   & \xmark    & \checkmark   \\  [1mm] \hline

\cite{r116}   & 2019 & \begin{tabular}[c]{@{}c@{}}PLS of URLLC\end{tabular} & \xmark    & \checkmark     & \xmark    & \xmark  & \checkmark     & \checkmark     \\  [1mm] \hline

\cite{pls87}   & 2019 & \begin{tabular}[c]{@{}c@{}}PLS of 5G networks\end{tabular} &\checkmark     & \xmark   & \checkmark   & \xmark     & \xmark   & \checkmark    \\ [1mm] \hline

\cite{pls73}  & 2020 & PLS of 5G networks  & \checkmark     & \xmark   & \checkmark   & \xmark     & \xmark    & \checkmark   \\  [1mm] \hline

\cite{r113}  & 2020 & \begin{tabular}[c]{@{}c@{}}PLS of URLLC using NOMA\end{tabular}   & \xmark   & \checkmark    & \xmark      & \xmark     & \xmark       & \checkmark    \\  [1mm] \hline

\cite{r114} & 2021 & \begin{tabular}[c]{@{}c@{}}PLS of URLLC using Jamming\end{tabular}     & \xmark    & \checkmark    & \checkmark & \xmark    & \xmark   & \checkmark   \\  [1mm]  \hline

\cite{r110}  & 2021 & PLS of 5G networks  & \checkmark    & \xmark   & \checkmark   & \xmark   & \xmark  & \checkmark \\  [1mm] \hline

\cite{r4}  & 2021 & \begin{tabular}[c]{@{}c@{}}PLS of URLLC\end{tabular}     & \xmark & \checkmark  & \checkmark   & \xmark    & \xmark & \checkmark   \\ [1mm]  \hline

\cite{pls2}  & 2022 & \begin{tabular}[c]{@{}c@{}}PLS of 5G Industry\end{tabular}    & \checkmark     & \xmark    & \checkmark           & \xmark     & \xmark    & \checkmark \\  \hline

\cite{r104}   & 2022 & \begin{tabular}[c]{@{}c@{}}PLS of URLLC\end{tabular}  & \checkmark  & \checkmark     & \checkmark   & \xmark   & \xmark & \xmark      \\ [1mm]  \hline

\cite{pls23}    & 2023 & \begin{tabular}[c]{@{}c@{}}Key-less PLS\end{tabular}    & \checkmark    & \xmark      & \xmark     & \checkmark  & \xmark    & \checkmark      \\  [1mm] \hline

\cite{r108}  & 2023 & PLS of 6G networks   & \xmark      & \checkmark     & \checkmark  & \xmark       & \checkmark    & \checkmark    \\ [1mm]  \hline

Our Survey & -    & \begin{tabular}[c]{@{}c@{}}PLS of URLLC\end{tabular}  & \checkmark & \checkmark & \checkmark   & \checkmark  & \checkmark    & \checkmark \\  [1mm] \hline

\end{tabular}
\vspace{-4 mm}
\end{center}
\end{table*}

\subsubsection{Survey and Tutorials on PLS of URLLC}

There are very limited research studies, surveys, and tutorials currently present that have analyzed the potential of PLS in providing secure communication for URLLC service. After extensive study, we find the survey works presented in \cite{r116} and \cite{r4}  provide some initial observations and analysis of the potential technologies in enhancing the security of the finite blocklength URLLC signal transmission from the physical layer perspective in 5G wireless networks. Specifically, the authors in \cite{r116} discussed some key PLS methodologies and performance evaluation metrics for finite blocklength URLLC while focusing particularly on the impact of channel state information (CSI) on its secrecy performance. The survey also identified the importance of evaluating metrics like secrecy gap and secrecy rate intervals in terms of the bit-error rate (BER) of the channel for accurate performance evaluation of PLS schemes for Finite blocklength communication. Similarly, the survey in \cite{r4}  discussed the PLS techniques adopted for short packet URLLC communication scenarios with different security strategies. The authors have provided a detailed analysis to differentiate the performance evaluation metrics under two categories such as, ergodic-based and outage-based, depending on the finite blocklength criteria for URLLC. 

Meanwhile, the survey work presented in \cite{r104} included an analysis of the PLS performance in various URLLC enabling technologies and methods to evaluate its performance. Moreover, the work also discussed the impact of mobility on the PLS performance of the URLLC service in the unmanned aerial vehicle (UAV) assisted communication scenario. Apart from the aforementioned studies, there is no other survey work that discussed the security enhancement techniques for 5G URLLC service and the underlying signal transmission strategies to the best of our knowledge. To impart a better understanding of the recent developments and compare the contributions of our work we provide a holistic summary of the related surveys and tutorials on PLS for 5G/6G wireless systems and URLLC service in Table \ref{tab2}, highlighting the major contributions and focused research areas.

\subsection{Motivation and Contributions}

After an extensive study, we can observe that the majority of existing surveys on PLS include the review of state-of-the-art research contributions for large blocklength codes, their use cases, enabling technologies, and resource allocation strategies in specific application scenarios of IoT \cite{r105}, D2D communication \cite{r110}, and relay-assisted communication \cite{r107}, \cite{r109}. We also identify that there are very limited survey works present that solely discuss the security of finite blocklength URLLC service. Most importantly, the scope of the existing surveys on PLS of URLLC found in \cite{r4}, \cite{r104}, and \cite{r116} are very limited and lack an in-depth analysis of the role of physical layer parameters like signal blocklength, pilot signal length, CSI estimation error, and dense network scenarios that can potentially affect the overall PLS performance. Apart from this, there is no such study that discusses comprehensively about the physical layer security of intelligent reflecting surfaces (IRS) aided URLLC, mission-critical URLLC use cases, and machine learning (ML) enabled security enhancement techniques while identifying the inherent challenges in achieving PLS for URLLC to the best of our knowledge. Moreover, the emerging security enhancements integrating advanced technologies like AI and blockchain for future HRLLC service in 6G have not been discussed in any survey work presented on 6G PLS.

Apart from this, relative study and observation expose the lack of focused work on establishing a structured and extensive survey of related research contributions presenting the PLS enhancements for mission-critical URLLC service. The requirement of secure data transmission is essential for finite blocklength URLLC signal, as it is the backbone of various emerging mission-critical applications in 5G as well as future 6G. Hence, an in-depth study on providing PLS for URLLC and a detailed analysis addressing the aforementioned challenges is hugely required. This motivated us to present a comprehensive and conceptual survey that can provide the relevant details of PLS enabling technologies to provide secure URLLC service in an easily understandable way.  

\begin{figure}[t]
	\centerline {\includegraphics[width=\linewidth, height=15.2 cm]{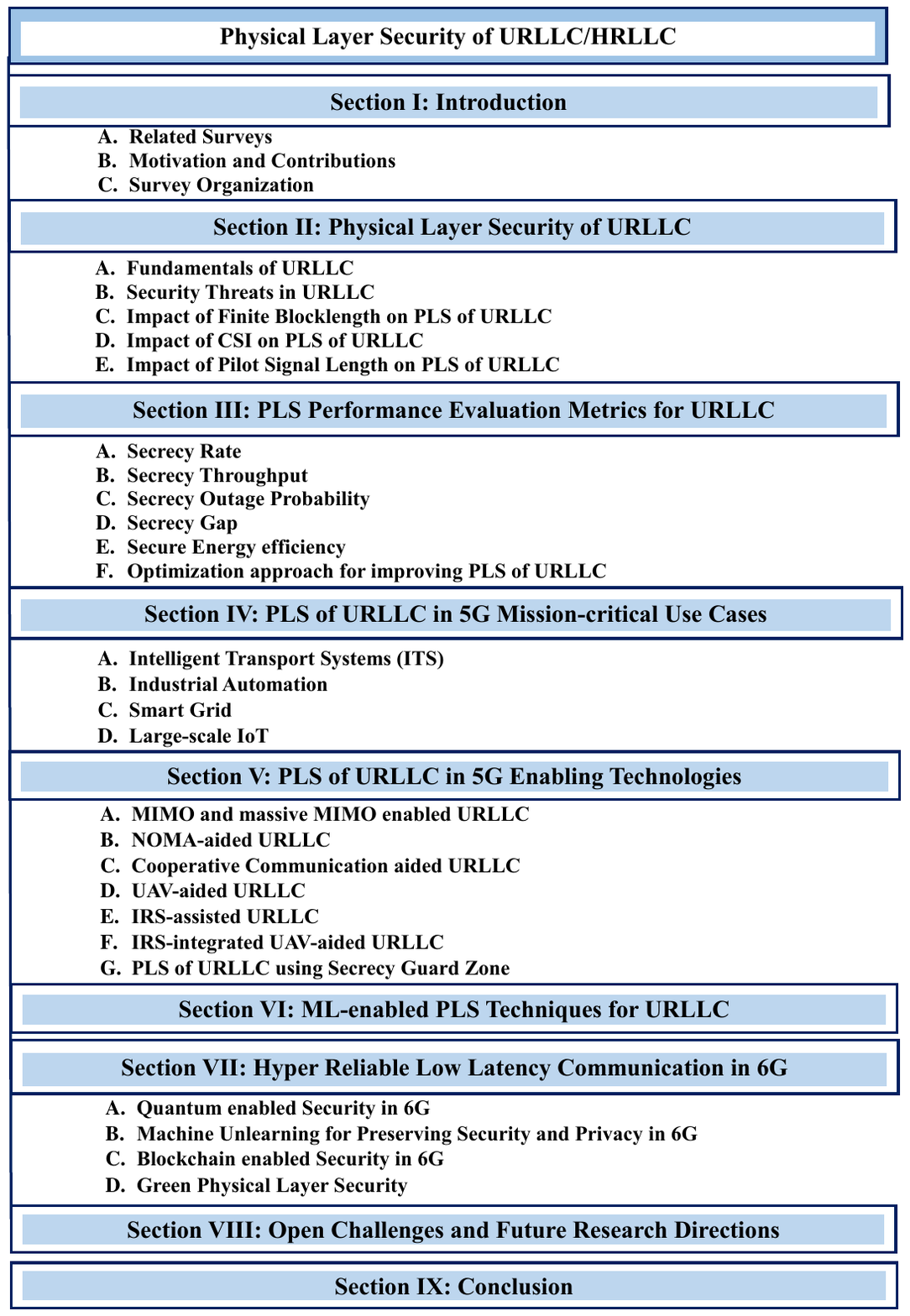}}
	\caption{The organizational structure of the survey}
	\label{fig1}
	\vspace{-5 mm}
\end{figure}

Thus, we conduct an extensive review of the state-of-the-art research studies and present this survey to provide the necessary knowledge regarding the PLS schemes used specifically for URLLC while identifying the inherent challenges and discussing the potential future research directions that can be explored for facilitating security, reliability, and low latency communication simultaneously in future 6G wireless networks. The contributions of this survey are summarized as follows:

\begin{itemize}

    \item This survey provides a comprehensive overview of the recent state-of-the-art research on the PLS enhancement technique for URLLC/HRLLC service in 5G/6G wireless networks. In particular, we provide a detailed justification of the requirement of PLS for ensuring secure communication instead of conventional key-based cryptographic techniques for finite blocklength URLLC by discussing the impact of various system design parameters on its performance. 
    
    \item We present a review of the impact of finite blocklength on the PLS of URLLC, including a detailed discussion on the specific PLS evaluation metrics like secrecy rate, secrecy gap, and secure energy efficiency used in various research works to measure its performance while considering the trade-off between the URLLC QoS and security. 
          
    \item Then, the survey includes an extensive and in-depth study of recent developments in PLS techniques used for various URLLC enabling technologies like NOMA, MIMO, massive MIMO (mMIMO), cooperating communication, and UAV and IRS relay-assisted communication.  
          
    \item  Moreover, we extend this survey to cover the advanced security enhancement schemes using ML-aided PLS for URLLC applications while discussing the use of distributed and decentralized ML-based security protocols for providing secure URLLC in this survey.
    
    \item Most importantly, our survey sheds light on the future security aspects in the 6G envisioned new service category HRLLC. We extensively discuss the role of various key emerging technologies like quantum communication and blockchain in providing robust PLS design for 6G HRLLC.
          
    \item  Finally, the survey identifies various open issues and challenges in developing efficient PLS strategies for URLLC/HRLLC while highlighting several promising future research directions to mitigate those challenges according to new requirements and constraints for 5G and future 6G applications.      
\end{itemize}

The survey will provide the necessary insights for current research on security enhancement of mission-critical applications supported by 5G URLLC and help researchers from both the industries and academia to gain relevant information on secure URLLC service under this one big comprehensive survey work. 
  

\subsection{Survey Organization}

The organization of the survey is represented in Fig. \ref{fig1}. First, in Section II, we provide a detailed analysis of the unique characteristics of finite blocklength coding-based URLLC service and its impact on the PLS design, including a discussion on various security threat models and the impact of system design parameters like CSI estimation and pilot-assisted communication. Then, Section III includes specific performance evaluation metrics adopted to evaluate the PLS performance of URLLC while achieving its QoS requirements simultaneously. In Section IV, a detailed study of PLS techniques utilized in various URLLC mission-critical application scenarios is presented. Furthermore, Section V includes an extensive survey of various novel techniques for ensuring PLS in URLLC along with a review of the state-of-the-art research works in MIMO, NOMA, cooperating communication, and UAV and IRS relay-assisted communication. Meanwhile, Section VI provides the recent developments in ML-enhanced PLS techniques for enabling URLLC service. In Section VII, a futuristic outlook of the unknown security threat model, its impact on the data transmission of HRLLC service, and emerging new technologies for facilitating intelligent and adaptive PLS for URLLC/HRLLC applications in 6G are discussed. Furthermore, Section VIII of the survey discusses various open research challenges and key potential future research directions for further research.  Finally, section IX provides the concluding remarks for the survey.

\section{Physical Layer Security of URLLC}

The early investigation of PLS is well illustrated in the information-theoretic analysis of secure communication given by Shannon in \cite{shannon1949communication}, which evaluates the secrecy of data transmission in terms of perfect secrecy depending on the amount of information leakage to the eavesdroppers. Generally, cryptographic techniques are adopted to prevent information leakage and establish secure communication in wireless systems. Security enhancement through cryptography is based on an encryption algorithm where the source node encrypts the data and shares a secret key with the legitimate receiver \cite{r30}. The receiver uses a decryption algorithm and the shared key to recover the original signal. Without the secret key, it becomes difficult for eavesdroppers to intercept the signal transmitted. However, eavesdroppers can still decrypt the data using an exhaustive key search. These conventional cryptographic algorithms introduce significant signaling overhead and processing delay due to multiple rounds of processing for encryption and decryption \cite{r22}, \cite{pls23}.  Moreover, cryptographic schemes depend on secure key management for secure communication. However, updating encryption keys can be difficult and time-consuming for URLLC applications due to dynamically changing wireless network conditions. Therefore, these conventional security schemes are not suitable for latency-sensitive URLLC applications. In such a scenario, PLS has emerged as a potential alternative technique to provide secure communication. It is beneficial for URLLC applications where data confidentiality and integrity are of utmost importance.  The information-theoretic secrecy analysis of PLS assures the security of URLLC without getting affected by the computational capability of the eavesdroppers in the network.

With the change in wireless technology, the implementation of PLS techniques has to deal with fundamental challenges like imperfect or partial CSI observation, influence of fading \cite{pls95}, secrecy coding design, and improvement of secrecy capacity in the presence of strong eavesdropping \cite{pls21}. Moreover, the dynamic nature of the wireless environment and stringent URLLC QoS requirements will introduce unprecedented challenges in terms of information leakage and security threats. Therefore, we provide a detailed analysis of the impact of finite blocklength coding and various system parameters like CSI and pilot signal length on the PLS performance of URLLC in the following sub-sections.

\subsection{Fundamentals of URLLC}

 To illustrate the security of URLLC, we need to revisit the fundamentals associated with this innovative service class of 5G. According to the 5G standards \cite{3gpp2017study}, \cite{pls76}, URLLC is a service type that enables high reliability in signal transmission while keeping the packet loss probability within $10^{-5}$ to $10^{-7}$ \cite{pls78}. URLLC in 5G brings significant advantages to various sectors and applications, offering higher reliability, faster response time with low latency communication ($\leq 1 ms$), and enabling transformative technologies that rely on real-time implementation in mission-critical applications to improve the overall user experience  \cite{r101}. However, the design parameters for URLLC packets need to address the trade-off between latency (measured in terms of coding blocklength) and packet error probability (that shows the reliability of data transmission), which enforces that the packet size should be small for URLLC. On the other hand, this short packet transmission incurs an inevitable non-zero decoding error, which degrades the reliability of the signal transmission.

 The conventional wireless communication system was specifically designed for infinite blocklength signals. Therefore, Shannon’s capacity becomes achievable for these systems while decreasing the decoding error probability to zero. Meanwhile, the payload size of URLLC signals is small with a finite blocklength \cite{pls97}. Due to the finite blocklength, conventional Shannon’s capacity-based asymptotic analysis does not apply to URLLC signals \cite{r29}. Therefore, the authors in \cite{polyanskiy2010channel} have presented an approximation for achieving the minimum achievable rate over additive white Gaussian noise (AWGN) channels for finite blocklength URLLC signals considering the signal blocklength, signal-to-noise ratio (SNR) and decoding error probability as: 
 
\begin{equation}
	R_{u} \approx  log_{2}(1+\gamma_{u})  - \sqrt{\frac{V(\gamma_{u} )}{N}}Q^{-1}(\epsilon),
 \label{eq1}
\end{equation}

where $R_{u}$ is the achievable rate of URLLC user $u$. Here, $\gamma_{u}$ is the SNR of the channel.  The channel dispersion parameter for URLLC user is denoted as $V\left ( \gamma_{u} \right ) = 1 - (1 + \gamma_{u} )^{-2}$. The inverse Gaussian Q-function $Q^{-1}(.)$ is denoted as $Q(z) = \frac{1}{2\pi }\int_{z}^{\infty }e^{-t^{2}/2} dt$. Here, it can be observed that the achievable data rate of the URLLC user $u$ in the finite blocklength regime is affected by back-off factors like $\epsilon$, which is the decoding error probability.

\subsection{ Security Threats in URLLC}

URLLC signal transmission can particularly get disturbed by some specific types of security attacks that affect the fundamental protocols of physical layer communication. From the physical layer perspective, security threats like jamming, eavesdropping, and pilot contamination attacks are highly responsible for degrading the URLLC signal quality \cite{pls21}. To illustrate more, we need to understand the nature of these attack types and the impact on PLS while relying on short packet communication.

\subsubsection{Jamming Attack}
 Jamming attack particularly involves some noise signals that are transmitted to disrupt the original signal transmission between the legitimate transmitters and receiver nodes \cite{r111}. To understand the impact of jamming, we need to understand the underlying concepts of wiretap channels as mentioned for the first time in Wyner's pioneering work in \cite{r1}. As per Wyner, the channel between the legitimate transmitter and the eavesdropper is called the wiretap channel. Wyner proposed that perfect secrecy can be achieved in the presence of eavesdroppers if the channel quality of the wiretap channel becomes worse than that of the legitimate channel. This is the fundamental reason behind the development of jamming, which is deliberately used to degrade the quality of the wiretap channel.

 Jamming has been considered as a technique of generating some undesired noise signal to disturb the original message transmission \cite{r114}, \cite{pls35}. This undesired noise signal is termed as artificial noise (AN), which helps to regulate original signal transmission while disrupting the information leakage to the eavesdroppers \cite{r112}, \cite{pls39}. The addition of AN with the confidential message improves the PLS performance for URLLC while actively countering the effect of strong eavesdropping. For instance, work in \cite{r30} has proposed a time-fraction-based URLLC signal transmission where a fraction of the time slot is utilized for confidential information and the remaining fraction for AN signal transmission.  Moreover, the authors in \cite{r22} and \cite{r57} have injected an AN signal into the null space of the channel vector of the legitimate link to ensure secure URLLC by negating the impact of eavesdropping.  For a more advanced scenario, the time series analysis of the jamming signal becomes quietly helpful in mitigating its impact while providing security to the URLLC signal transmission \cite{pls36}.
 	
\begin{figure}[t]
	\centerline {\includegraphics[width=1\linewidth, height=5 cm]{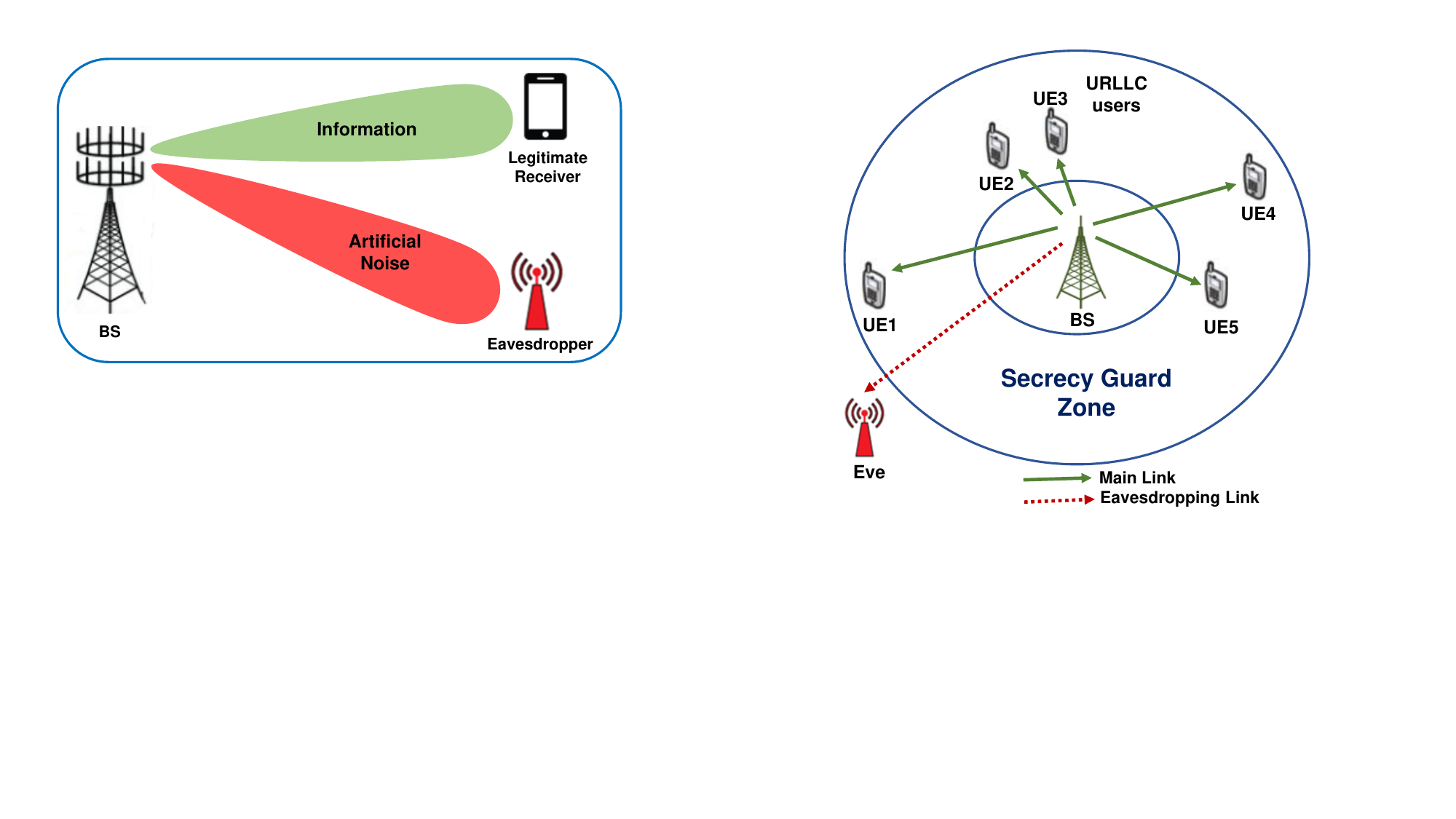}}
	\caption{Jamming using Artificial Noise signal}
	\label{fig2}
	\vspace{-4 mm}
\end{figure}

\subsubsection{Eavesdropping}
Eavesdropping is a process of intentionally interfering or wiretapping the wireless channel to intercept the confidential message signal. Depending on the mode of eavesdropping, they can be classified as active or passive eavesdroppers. These eavesdroppers can present near legitimate users as internal eavesdroppers or present externally to carry out the security attack. Basically, the active eavesdropper, which is otherwise termed as colluding eavesdropper directly interferes with the original message signal transmission to degrade the received signal quality at the legitimate receiver. 

On the other hand, passive or non-colluding eavesdroppers intercept the confidential information exchange exchanged between the legitimate transmitter and receiver without degrading its quality. Generally, active eavesdroppers inject AN or jamming signals directly to the legitimate channel if they exist in the beam space of the transmitter. However, passive eavesdroppers operate in passive mode and do not transmit AN or jamming signals. As a result, acquiring the location information and the instantaneous CSI of these eavesdroppers is challenging and impractical \cite {r22}. Therefore, many researchers utilized the statistical CSI of the legitimate channel and the eavesdropping channel to design an appropriate PLS scheme for URLLC.

\subsubsection{pilot contamination attack}
 Another security attack affecting URLLC is the pilot contamination attack (PCA). In PCA, the eavesdropper intentionally transmits identical pilot signals to disrupt the channel estimation process and cause information leakage. For URLLC, the pilot signal length is very small to adhere to the latency constraint, which increases the chance of contamination. To prevent PCA, enlarging the received signal strength at the legitimate receiver over the eavesdropping channel by maximizing the secrecy rate is an efficient technique.

 Eavesdroppers can alter the channel estimation results by performing a PCA directly or with the help of relays (i.e., UAV or IRS) in short packet transmission. But, in the case of URLLC, the pilot sequence length is deliberately made short to adhere to the low latency constraint. This does not allow the use of random modulation or adding random bits to the pilot sequence for possibly countering the PCA effect. Therefore, the detection of PCA and ensuring secure data transmission is one of the major challenges and requires further study to ensure secure URLLC signal transmission.

\subsection{Impact of Finite Blocklength on PLS of URLLC}

Apart from reliability and latency, the security of the finite blocklength URLLC signal is one of the major concerns for various mission-critical URLLC applications. The low- latency of URLLC limits the codeword size of each message signal into relatively smaller blocklength after encoding \cite{r9}.  As a result, a bust of small payload (i.e., only hundreds of bits) of URLLC data is generated and transmitted, which is evidently different than the usual long codeword signal transmission. This implies that the conventional Shannon capacity-based analysis will not be able to characterize the achievable rate of short packet finite blocklength signal. Meanwhile, the widely used secrecy capacity-based performance evaluation method defined for larger blocklength signals will become ineffective for URLLC. This is because the conventional secrecy capacity is only achievable if the signal block length asymptotically increases to infinite. Therefore, the achievable secrecy rate of finite blocklength can be approximated in terms of decoding error and information leakage as \cite{r10},
\begin{equation}
	R_{s} \approx  C_{s}  - \sqrt{\frac{V(\gamma_{u} )}{N_{u}}}Q^{-1}(\epsilon )- \sqrt{\frac{V(\gamma_{e} )}{N_{u}}}Q^{-1}(\delta). 
\label{eq2}
\end{equation}
 The above expression of the secure data rate $R_{s}$ is widely used and valid for finite blocklength signals with the channel condition, where $\gamma_{k}\geq \gamma_{e}$, and becomes zero otherwise where $\gamma_{u}$ and $\gamma_{e}$ are the SNR of the legitimate channel and the eavesdropping channel respectively.  Here, $C_{s}$ is the secure channel capacity and can be expressed as $C_{s}= log_{2}(1+\gamma_{u})-log_{2}(1+\gamma_{e})$.  We can observe that the achievable secure data rate of user $u$ in the finite blocklength regime is affected by back-off factors like $\epsilon$, which is the decoding error probability at the legitimate receiver and the information leakage probability of the eavesdropper $\delta$.  Here, the channel dispersion parameter is denoted  as $V\left ( \gamma_{w} \right ) = 1 - (1 + \gamma_{w} )^{-2}$ for $w \in \left \{ k,e \right \}$. Channel dispersion at high SNR can be approximated to 1, but for low SNR conditions, this approximation does not hold well \cite{r26}. Here, Eq. \ref{eq2} represents the dependence of the secure transmission rate on the decoding error and information leakage in the finite blocklength regime. From the information-theoretic perspective, as long as the coding rate of URLLC does not exceed $R_{s}$ with a confidential codeword length $N_{u}$ after encoding, both the reliability constraint $\epsilon$ and secrecy constraint $\delta$ can be guaranteed.

\subsection{Impact of CSI on PLS of URLLC}

In order to have secure communication between the legitimate transmitter and receiver, knowledge regarding the instantaneous CSI of the legitimate receiver and the eavesdropper becomes crucial. This helps the transmitter to adjust its transmit power based on the instantaneous CSI of the attacker and establish secure communication. However, acquiring instantaneous CSI for ensuring PLS of URLLC is challenging due to the dynamic wireless environment and delay in getting feedback \cite{r22}, \cite{r153}. Therefore, the prior knowledge of CSI of the legitimate channel and the eavesdropping channel is also essential for the legitimate transmitter to design efficient and secure transmission strategies. So, keeping this in mind, the optimal system design and performance evaluations of PLS for URLLC can be done through the proper channel training for obtaining CSI values \cite{r107}. However, the possibility of channel estimation error and delayed feedback is a bottleneck for ensuring the PLS of URLLC in practical scenarios. 

In many situations, accurate CSI estimation is possible for legitimate channel, whereas it becomes difficult for eavesdropping channel due to the passive and hostile nature of the eavesdroppers. Therefore, only statistical CSI can be obtained based on past channel observations for eavesdroppers \cite{r101}. This leads to the assumption of having perfect CSI for evaluating the secrecy performance of PLS through secrecy rate or secrecy throughput in case of finite blocklength communication under ideal conditions. However, the finite blocklength of the URLLC signal imposes challenges for accurate CSI estimations of the channel. Firstly, due to the small blocklength and strict latency constraint, there is a lack of enough channel uses for channel estimation and feedback reception. Furthermore, accurate CSI estimation may incur pilot overhead and consume more power, which may not be feasible for massive IoT systems using short-packet communication \cite{r105}. Thus, requires pilot length optimization for finite block length URLLC signal while designing secure transmission policies through pilot-based channel training.

\subsection{Impact of Pilot signal length on PLS of URLLC}

PLS techniques primarily focus on wireless channel characteristics to provide secure communication. Therefore, accurate CSI estimation is essential to ensure PLS for URLLC types of services. The prior knowledge of instantaneous CSI is indispensable for secure and reliable signal transmission \cite{r60}. To increase the robustness of short packet communication (SPC) and improve reliability, pilot-based signal transmission is extensively used \cite{r17}, \cite{r25}, \cite{r19}, \cite{r49}. Specifically, pilot-assisted communication is adopted for CSI estimation, which is the key requirement for reliable data transmission. However, for URLLC, the pilot signal length is optimally chosen to address its low latency constraint. 

Large pilot symbol length improves CSI estimation accuracy but reduces the message signal length in a given blocklength at which the URLLC is transmitted. On the contrary, reducing pilot signal length may not provide sufficient time to obtain accurate CSI and degrade the received signal quality and reliability \cite{r18}. The study given in \cite{r57} has presented a framework for the design of appropriate PLS schemes in the pilot-assisted URLLC signal transmission. Here, the authors have emphasized the requirement of optimizing the pilot length and blocklength simultaneously to improve the secrecy performance of the system. This suggests that the optimal pilot length selection for finite blocklength URLLC signals is necessary to improve system performance. Moreover, in the case of URLLC-MIMO systems, pilot-assisted channel training has optimized the beamforming, which ultimately improves the security of the received signal \cite{r40},\cite{r43}. Therefore, in practical systems, designing proper channel training to obtain accurate CSI for improving the security of URLLC becomes crucial from the physical layer perspective.

\subsection{Summary and Insights}

In summary, from the information-theoretic perspective of PLS, we can say that even for a low achievable secrecy rate of URLLC, decoding error, and information leakage are inevitable.
There is an interesting relation between the blocklength and secrecy rate. With the increase in blocklength, decoding error diminishes which ultimately improves the achievable secrecy rate. However, increasing signal blocklength leads to an increase in overall latency. This implies that there exists a trade-off between the reliability, latency, and security constraints of URLLC which need to be handled carefully while developing PLS schemes.

The broadcast nature of wireless channels makes wireless communication vulnerable to security threats. Many mission-critical 5G applications using short packet communication are facing challenges in providing secure communication. The increased network traffic, strict latency, and reliability requirements make it even more difficult to ensure PLS. Specifically, in large-scale heterogeneous networks, providing PLS needs security enhancement techniques that can satisfy 5G service requirements and secure communication simultaneously. Therefore, security enhancement techniques optimizing pilot signal length, URLLC blocklength, and accurate CSI estimation specific to URLLC constraints are highly desirable for secure data transmission in 5G URLLC applications.

\section{PLS performance evaluation metrics for URLLC}

Evaluation of PLS performance for secure URLLC is remarkably critical from the transmission design perspective. Conventionally, secrecy capacity has been used as a prime secrecy metric to evaluate PLS performance for long blocklength signals \cite{pls43}. But, for URLLC, it cannot be used as a secrecy metric where the signal blocklength is finite, and it incurs a non-negligible error probability at the receiver \cite{r7}. Therefore, the legitimate URLLC user channel capacity $C_{u}$ and the eavesdropping channel capacity $C_{e}$ cannot eliminate the decoding error probability ($\epsilon$), which ultimately affects secure transmission.  Similarly, the secrecy metrics like secrecy outage probability (SOP), connection outage probability (COP), and perfect secrecy are not effective enough to evaluate the PLS performance for URLLC because these metrics are defined specifically infinite blocklength signals, whereas URLLC uses finite blocklength signals for communication \cite{r54}. This demonstrates the necessity of fair secrecy metrics to evaluate the PLS performance considering URLLC transmission rate $R_{u}$, decoding error probability $\epsilon$, SNR $\gamma_{u}$, and finite blocklength of size $N_{u}$.

In this regard, various research works have utilized the aforementioned system parameters to develop the secrecy performance evaluation metrics specific to URLLC, considering the strict reliability and latency bounds. Hence, the PLS performance metrics that take into account finite blocklength criteria can be evaluated in terms of secrecy rate, secrecy throughput, secrecy gap, secrecy outage probability, and secure energy efficiency \cite{r6}, \cite{r9},\cite{r10}. In various URLLC application scenarios, these secrecy evaluation parameters are regarded as the goal of the optimization problems. Therefore, a brief summary of the state-of-the-art contributions that include various secrecy performance evaluation metrics for URLLC like secrecy rate, secrecy throughput, secrecy gap, and secrecy outage probability (SOP) along with the optimization approaches used for improving the overall PLS performance is given in the following sub-sections.  

\subsection{Secrecy Rate}
In PLS, the secrecy rate is defined as the number of secure bits transmitted per channel use \cite{r11}. For URLLC, the blocklength is finite and is denoted as $N_{u}$. Therefore, the minimum number of channel uses required for transmission of URLLC is also equal to $N_{u}$. Assuming $B_{u}$ as the number of information bits transmitted over $N_{u}$ channel uses, the maximum coding rate becomes $R= B_{u}/N_{u}$ \cite{r61}.  Therefore, the secrecy rate $R_{s}$ can be described as the difference between the data rate of legitimate URLLC user $R_{u}$ and the eavesdropper $R_{e}$, i.e., $R_{s}$ = $R_{u}$-$R_{e}$ \cite{r11}. Here, the successful secure data transmission can only be possible when $R_{u} \leq C_{u}$ and $R_{e} \geq C_{e}$. However, perfect secrecy may not be achievable for finite blocklength signals due to the presence of non-zero decoding error probability. As the ergodic capacity metrics are not suitable for finite blocklength secrecy analysis, the effective secrecy rate can be evaluated by jointly optimizing key parameters like pilot signal length and power allocation for URLLC signals \cite{r4}.

\begin{figure}[t]
	\centerline{\includegraphics[width=0.95\linewidth]{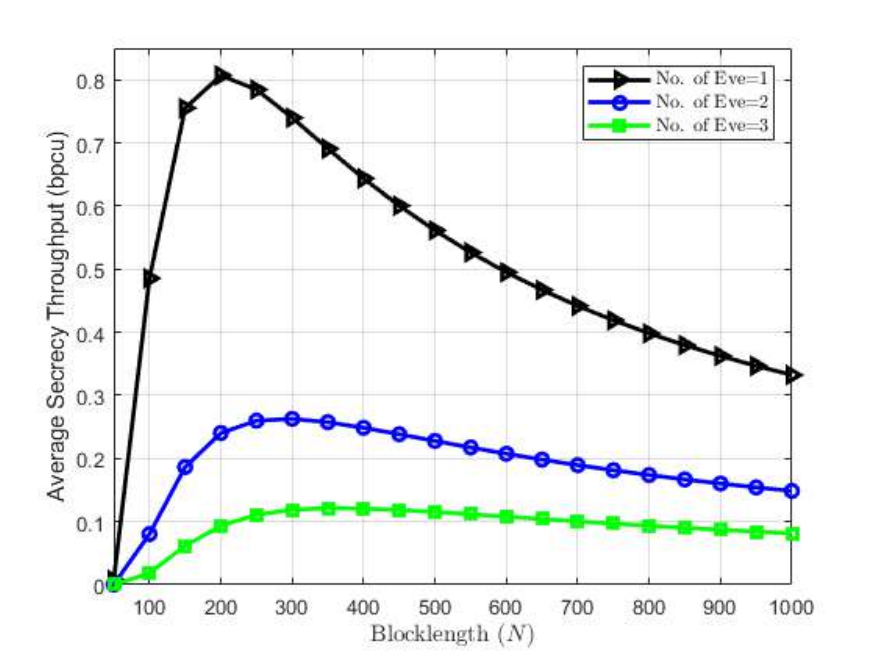}}
	\caption{Average Secrecy Throughput vs. Blocklength $(N)$. }
	\label{fig3}
 \vspace{-4mm}
\end{figure}

\begin{figure}[t]
	\centerline{\includegraphics[width=0.95\linewidth]{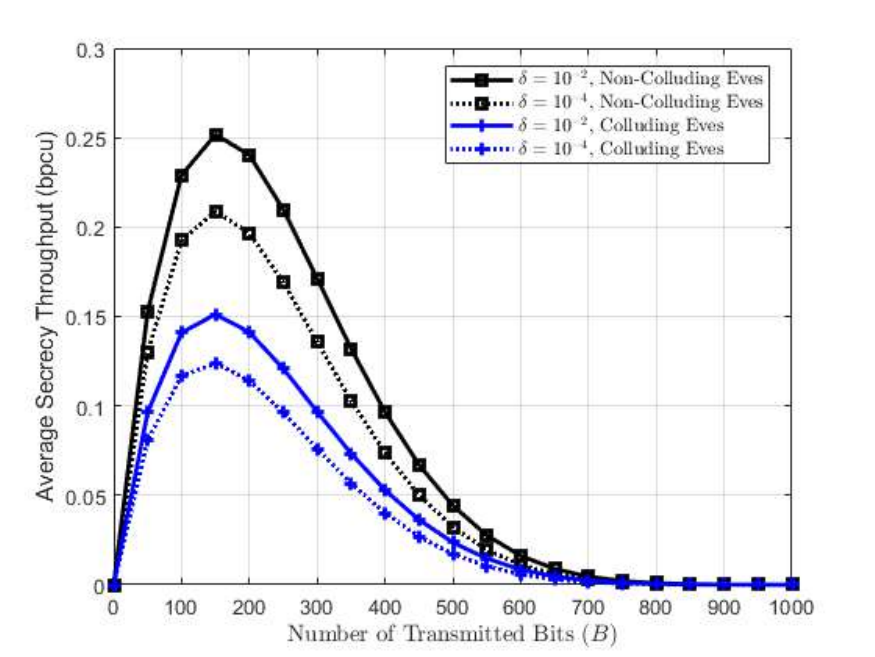}}
	\caption{Average Secrecy Throughput vs. Transmitted information bits. }
	\label{fig4}
 \vspace{-4mm}
\end{figure} 

\subsection{Secrecy Throughput}
For finite blocklength URLLC signal, secrecy throughput evaluation can be determined as the average signal transmission rate under a certain reliability and secrecy constraint \cite{r23}. Secrecy throughput is the most commonly used PLS evaluation metric for URLLC as it captures the system performance under URLLC constraints and predefined secrecy criteria. The analytic framework given in studies like \cite{r56}, \cite{r22}, \cite{r52}, and \cite{r15} provides the closed-form approximation of secrecy throughput for URLLC signals. Additionally, the authors of the aforementioned studies discussed the impact of the system. parameters like signal blocklength, intercept probability, and type of eavesdropping on the QoS of URLLC and security.

We can observe the variation of average secrecy throughput with respect to signal blocklength, number of eavesdroppers, transmitted information bits, and type of eavesdropping in Fig. \ref{fig3} and Fig.\ref{fig4}. Specifically, Fig. \ref{fig3} shows the initial increase in secrecy throughput with respect to blocklength, Then, after a certain blocklength value, it starts decreasing. This happens because, with the increase in blocklength, the secrecy rate of URLLC transmission increases.  However, the corresponding decoding error probability increases with the increase in data rate. This ultimately affects the overall secrecy throughput performance. Similarly, the number of eavesdroppers, secrecy constraints, and type of eavesdroppers adversely affect the secrecy throughput of finite blocklength URLLC as shown in Fig. \ref{fig4}.

\subsection{Secrecy Outage Probability}
Apart from secrecy rate and secrecy throughput, secrecy outage probability (SOP) is another important performance evaluation metric for wireless communication \cite{pls40}. However, measuring SOP as defined for conventional infinite blocklength signals is not efficient enough to capture the inherent impact of non-zero decoding error probability of finite blocklength URLLC signal \cite{r8}, \cite{r57}. Therefore, the authors in \cite{r57} have defined a modified SOP metric considering the impact of decoding error probability to evaluate the PLS performance of pilot-assisted URLLC service. Here, the definition of SOP for URLLC signals is different from the conventional large blocklength signals. The SOP for URLLC is defined as the probability of not satisfying either the reliability constraint (measured in terms of decoding error probability) or the secrecy constraint (measured in terms of intercept probability) during secure communication of URLLC signals with a target data rate. Furthermore, the defined SOP metric reflects the PLS requirement for finite blocklength signals and balances the reliability and security trade-off for URLLC \cite{r14}, \cite{r28}.

\subsection{Secrecy Gap}
Moving from information-theoretic analysis, the authors in \cite{r4}, have proposed a secrecy metric based on the average bit-error rate called secrecy gap (SG). SG is basically defined as the ratio between the minimum SNR threshold value of the legitimate channel and the maximum SNR value of the wiretap channel. This represents a balance between the reliability and security constraints of URLLC. Moreover, SG demonstrates that the average bit-error-rate (BER) of the main channel should not exceed the maximum BER condition considering the finite blocklength signal. On the other hand, secure communication will only be established if the average BER of the eavesdropper channel exceeds the minimum BER value. This makes SG suitable for evaluating PLS performance of URLLC. 

Similarly, another metric to evaluate PLS can be the measurement of secure rate interval, which is the difference between the highest allowable transmission rate of the legitimate user to the lower bound of the transmission rate satisfying the URLLC criteria. This implies that eavesdroppers cannot intercept confidential information as long as the eavesdropping rate $R_{e}$ value is greater than eavesdropper channel capacity $C_{e}$. Hence, careful evaluation of PLS performance using SG and secure rate interval is essential to improve the performance of secure communication for URLLC signal transmission in 5G applications.

\subsection{Secure Energy efficiency}

Ensuring energy efficiency and security of URLLC signal transmission in energy-constraint applications like large-scale IoT networks and UAV-aided communication is highly desired \cite{r17}. In wireless communication, secure energy efficiency (SEE) refers to the amount of energy consumed to maintain reliable communication under specific security constraints \cite{r47}. SEE can be defined as the ratio between the secure throughput and the total power consumption of the system \cite{pls52}, \cite{r2}. It is the measure that represents efficient communication while safeguarding against potential security threats. This requires designing efficient PLS schemes that minimize energy consumption without compromising the security of signal transmission. SEE can be achieved by combining PLS schemes with strategies like adaptive modulation, efficient resource allocation, and optimized power allocation techniques under the constraints of URLLC QoS and security \cite{pls56}.

\begin{table}[]
\caption{Summary of PLS metrics used  for URLLC in related research contributions}
\label{tab3}
\begin{tabular}{|l|p{18em}|}
\hline

\multicolumn{1}{|c|}{\textbf{ PLS Evaluation Metric}} & \textbf{\begin{tabular}[c]{@{}c@{}} References \end{tabular}} \\ \hline


Secrecy Rate  & \begin{tabular}[c]{@{}l@{}} \cite{r61},\cite{r9}, \cite{r12} \end{tabular} \\ \hline

Secrecy Throughput  &  \begin{tabular}[c]{@{}l@{}} \cite{r10},\cite{r11},\cite{r14},\cite{r56}, \cite{r22}, \cite{r52}, \cite{r15}, \cite{pls83}, \cite{uav8} \end{tabular} \\ \hline
 
Secure Outage Probability  &  \begin{tabular}[c]{@{}l@{}} \cite{r57}, \cite{r28} \end{tabular} \\ \hline

Secrecy Gap  &  \begin{tabular}[c]{@{}l@{}} \cite{pls11}, \cite{r4}, \cite{pls49}\end{tabular} \\ \hline

Secure Energy efficiency  &  \begin{tabular}[c]{@{}l@{}} \cite{r2},\cite{pls52}, \cite{r47} \end{tabular} \\ \hline

\end{tabular}
\vspace{-4mm}
\end{table}

\subsection{Optimization approach for improving PLS of URLLC}

In this subsection, we focus on fundamental optimization problems and solution methodologies adopted to ensure the security of URLLC from the physical layer perspective. Specifically, the optimization methods presented in the state-of-the-art research works suggest maximization of achievable secrecy rate and secrecy throughput, minimization of SOP, and maximization of energy efficiency under specific security and URLLC QoS constraints. To achieve these objectives, various optimization frameworks are adopted, which we summarize as follows. 

\subsubsection{Resource Allocation optimization}

Optimizing the physical layer resources for improving the PLS of URLLC is considered as a promising technique which has been investigated by many researchers recently. Designing secure resource allocation involves optimization of specific parameters like URLLC signal blocklength, decoding error probability, sub-carrier allocation, and secrecy constraint. For URLLC, optimizing these parameters leads to the maximization of achievable secrecy throughput and minimization of SOP. Generally, the formulated optimization problem for improving PLS performance is a mixed integer programming problem \cite{pls52}, \cite{r9}. This is due to the presence of integer constraints like blocklength of URLLC signals \cite{r16}. In \cite{r3}, the authors jointly optimize the blocklength, transmit power, and the number of information bits in the finite blocklength code to maximize the secure spectral efficiency of the URLLC system. Similarly, in \cite{r29}, maximization of the minimum secrecy rate of short packet communication (SPC) has been achieved by jointly optimizing the transmit power, AN power, and the time fraction for transmitting the information and AN signal.

\subsubsection{Beamforming and Precoding optimization}
PLS improvement through secure beamforming and precoding is applied extensively for URLLC in the multi-antenna deployment scenario, or MIMO systems. Basically, the transmitter uses the beamforming technique to concentrate the transmit power towards the desired receiver \cite{r114}. Conventional beamforming techniques like zero-forcing (ZF) and maximum ratio transmission (MRT) are often used to control the beam pattern and direct it toward the legitimate receiver while suppressing it toward the eavesdropper. It can be observed that semi-definite relaxation (SDR) and alternating optimization methods are extensively used for designing secure beamforming and to solve optimal power allocation problem \cite{r75}, \cite{r76}. However, optimal beamforming design depends on the availability of eavesdroppers CSI. Furthermore, the beamforming and precoding rely heavily on CSI. However, massive MIMO and mmWave communication complicates the CSI estimation process and, in return, decreases the efficiency of the secure technique. Moreover, optimizing the precoding technique has to deal with the trade-off between the design complexity and PLS performance. Therefore, the transmission of AN null space beamforming is widely adopted to disrupt the information leakage to the eavesdroppers and provide secure data reception at the legitimate URLLC user when the full knowledge of the eavesdropper's CSI is unavailable. This implies that to satisfy the desired secrecy level and maximize the secure throughput of URLLC, a careful design and optimization of the beamforming vector and precoding matrix is required.

\subsection{Summary and Insights}
The definition of perfect secrecy based on Shannon's work is the channel capacity of the legitimate receiver for a code length that tends to infinity. This perfect secrecy is also termed as strong secrecy. On the other hand, the concept of perfect secrecy does not exist for finite blocklength signals as there is a chance of outage associated due to the small blocklength \cite{pls1}. In this regard, many researchers have proposed efficient PLS techniques for the FBL regime based on Wyner's pioneering work \cite{r1}. His work provides the necessary foundation for secure communication and suggests that the achievable secrecy rate or perfect secrecy can be guaranteed over a wiretap channel for infinite blocklength signals.  However, for URLLC with finite blocklength, error-free transmission is not possible due to non-zero decoding error probability at the receiver. So, the application of conventional PLS methods for finite blocklength signals will provide a sub-optimal solution. 

Despite the advantages associated with the traditional secrecy metrics used for infinite blocklength signals, it fails in evaluating and characterizing the PLS performance for finite blocklength URLLC without considering the QoS requirements associated with it. Therefore, this section essentially emphasized on the discussion regarding the characteristic difference between the proposed secrecy metrics defined for URLLC in various research studies. Moreover, the keyless PLS metrics like secrecy rate, secrecy throughput, secrecy gap (based on SNR), secure energy efficiency, and modified secrecy outage probability provide a fair evaluation of PLS performance for finite blocklength URLLC. Another important aspect of PLS is to focus on the impact of various URLLC design parameters like decoding error probability, blocklength, imperfect or partial CSI, and SNR of both the legitimate and eavesdropping channels on its performance. Sometimes for improving the efficiency of the PLS scheme, the QoS requirements like throughput, and energy efficiency are compromised. Therefore, the PLS performance evaluation metrics for URLLC need to be conditioned accordingly to provide secure communication and QoS requirements simultaneously.


\section{PLS of URLLC in 5G mission-critical use cases}

In the past decade, the introduction of URLLC service has transformed the operation of various mission-critical applications like industrial automation, intelligent transport systems (ITS), and smart grids to a more automated and self-reliant one. Most importantly, the reliable and low latency signal transmission enables the automation process in these applications by reducing the margin of error.  It is well evidenced that the current development of these applications needs to fulfill the demand for stringent QoS requirements like high data rate, high reliability, improved connectivity, and low latency communication by 5G services. Additionally, these applications involve a large number of machine-type devices and end users which constantly communicate by transmitting confidential data and control information (i.e., URLLC signal) among them. This makes the signals vulnerable to security threats and eavesdropping attacks. In this regard, several research studies focus on developing efficient PLS schemes for URLLC by exploiting resources like optimal power allocation, beamforming design, and physical channel parameters in mission-critical 5G applications. Hence, to gain the required insights, we briefly discuss the security threats affecting URLLC signal transmission and the potential PLS enhancements proposed in the related studies for enabling secure URLLC in mission-critical use cases like ITS, industrial automation, and smart grid applications in the following sub-sections.

\begin{figure*}[t]
	\centering
	\includegraphics[width=\linewidth, height=9.8 cm]{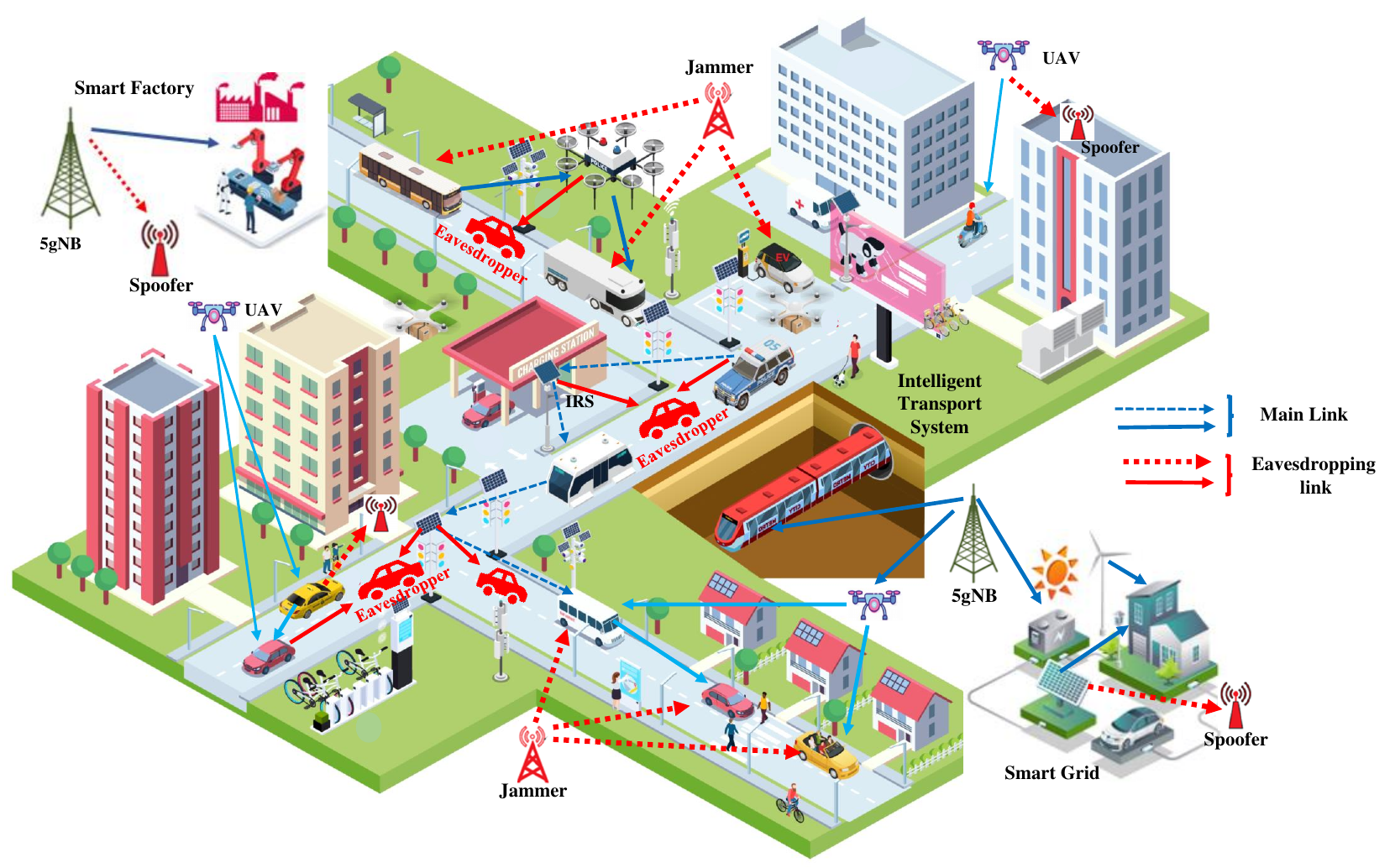}
	\caption{PLS of URLLC in 5G mission-critical use cases}
	\label{fig5}
	\vspace{-4mm}
\end{figure*}

\subsection{Intelligent Transport Systems (ITS)} 

The introduction of URLLC has revolutionized the current ITS applications like autonomous driving and vehicle-to-everything (V2X) by supporting extreme QoS requirements. Recently, vehicular communication has opened up new opportunities for improving road safety and efficient routing in autonomous driving using 5G services. Additionally, V2X communication introduces remarkable advantages to ITS by efficiently transmitting delay-sensitive data, multimedia streams, and mission-critical URLLC information among vehicles, roadside units, sensor nodes, and cellular base stations \cite{pls69}. This allows vehicular users to exchange real-time information about traffic, road conditions, routing data, location and speed of vehicles, breaking status, and the data generated from sensors through a wireless medium. However, broadcasting this delay-sensitive URLLC control information through the open wireless medium is prone to security attacks and eavesdropping. Therefore, preserving the confidentiality and security of this information is vital to avoid unwanted accidents and security breaches.

 In such circumstances, key-based or cryptographic techniques will not be a good fit as they increase the complexity of computation. So, PLS schemes are the perfect match for secure URLLC service in ITS. However, there are some limiting factors like the mobility of vehicles \cite{pls85}, limited radio resources, and delay-sensitive applications, which incur challenges for PLS design \cite{r123}, \cite{pls69}. Many PLS-based studies on ITS addressed the security attacks from colluding eavesdroppers \cite{r72}, adversary jamming attacks \cite{r114}, emulation attacks, and passive eavesdropping \cite{pls70}. Mitigating such security threats can be possible by integrating other techniques like cooperative, friendly jamming, NOMA, and physical layer authentication. This can potentially help to achieve a positive secrecy rate for URLLC signal transmission. The research studies presented in \cite{pls26} and \cite{pls29} have identified the key challenges in ensuring secure V2X communication for ITS. These studies have discussed some potential PLS techniques, which can be referred to continue further research. However, developing efficient PLS schemes for secure URLLC signal transmission in V2X communication is still in its infancy and demands further in-depth analysis considering URLLC QoS requirements, security constraints, and physical layer channel parameters.

\subsection{Industrial Automation}

The current revolution in industries is unprecedented, and it is only possible due to the introduction of 5G wireless communication, which is the key enabler of various mission-critical operations on the factory plane like advanced sensing techniques, automation, high precision operations, and edge computing \cite{pls2}. Moreover, the inclusion of service-oriented technologies, low-latency operations, and AI-based solutions is advancing the industrial automation process and improving productivity. The recently standardized 5G NR Release-16 and release-17 include various enhancements to improve the performance of industries \cite{pls64}.  Most importantly, the exponential growth of IoT-enabled industrial infrastructure, i.e., IIoT, is the backbone of the new industrial revolution. Specifically, the critical control information. exchange between the controller and actuator nodes requires secure, reliable, and low-latency signal transmission to support safe and efficient process control and operation. However, the presence of large-scale IoT devices, sensor nodes, actuators, robotics arms, and automated guided vehicles in a smart industry scenario makes secure data transmission a challenging task. 

 As observed, the industrial wireless network mainly suffers from security threats at different communication layers in terms of eavesdropping, jamming, and spoofing attacks \cite{pls1}. Therefore, as an essential countermeasure, PLS techniques, as proposed in studies like \cite{pls26} and \cite{r72}, provide lightweight security solutions without using complex cryptographic security schemes in industries. 

 Similarly, the work in \cite{r22}, \cite{r61}, and \cite{pls52} consider the industrial IoT (IIoT) scenario to provide efficient PLS schemes. In \cite{r22}, the confidential URLLC signal is integrated with AN signal by a multi-antenna controller for secure data transmission to the single-antenna actuators in the mission-critical remote control operations in the presence of distributed non-colluding eavesdroppers. The work in \cite{r61}, proposed a discriminatory channel estimation protocol combined with AN injection to disrupt the channel estimation capability of the eavesdroppers in a URLLC-IIoT system. Both the aforementioned works emphasized on improving the achievable effective secrecy rate of URLLC users. Further, the work in \cite{pls59} enhanced the PLS for URLLC signal by considering a static IIoT scenario where the actuators and controller are fixed in their positions without having mobility. The IIoT system contains a large number of low power IoT devices \cite{pls57}. Hence, providing energy-efficient security solutions is desired for efficient management of control and automation processes in IIoT. Apart from the above studies, the work in \cite{pls52} proposed an optimization framework for URLLC blocklength and transmit power to provide energy-efficient and secure data transmission in IIoT. 

\subsection{Smart Grid}

Smart Grid systems are becoming an integral part of the smart city infrastructure. The integration of automation and interconnection between massive sensor nodes through delay-sensitive data communication has transformed the conventional complex electricity grid operation into a more advanced smart grid system. Basically, the physical layer operations of smart grids like electricity generation, transmission and distribution, smart metering, and advanced communication infrastructure ensure uninterrupted power supply and enable quick fault detection and recovery processes.

Smart grids are often termed as smart energy systems because they can manage the traditional electric grid by enabling digitization through a data communication layer \cite{pls25}. The smart grid operations mainly use short packet URLLC signals to provide faster information exchange among the multiple sensors and actuator nodes. Reliable and low-latency information exchange of the real-time monitoring and control information, grid supervision, and fault detection data in URLLC ensures the stability of smart-grid operations \cite{pls55}. 

However, ensuring secure communication among the digital components, sensors, and remote radio units is challenging due to physical layer attacks like jamming and false injection of signals into the critical information flow of URLLC service. Thus, physical layer security concerns for URLLC in smart grid operations increase with the growing inclusion of machine-type devices, which enables the integration of wireless information exchange and digitization of processes. This results in information leakage and vulnerabilities at the device level that result from incorrect measurements, data fabrication, and man-in-the-middle attacks \cite{pls29}. In return, these factors hugely influence the URLLC data transmission among remote sensors, controllers, and measurement units. Similarly, false message injection, jamming, and the presence of unauthorized eavesdroppers at the data communication layer of a smart grid wireless local area networks can potentially lead to loss of energy at specific modes \cite{pls24}. There are some preliminary works present that explore the security requirements and challenges of smart grids, as in \cite{pls25} and \cite{pls24}. However, the focused research work toward providing security to URLLC data transmission is still in its initial stages. In such scenarios, PLS enhancement through optimized precoding, adding AN, friendly jamming, and ML-based security enhancement techniques can provide promising solutions for ensuring secure communication of URLLC signals. 

\subsection{Large-scale IoT}

The emergence of IoT has provided a paradigm shift to many futuristic applications by connecting everyone with everything. IoT is the basic backbone of emerging industrial automation and smart city applications. The unique capability of connecting billions of devices wirelessly for information exchange makes IoT applications indispensable. The IoT devices are basically operated with low power and communicate with each other by exchanging important and private information, such as control signal transmission in an industrial automation scenario among the controller and the actuators \cite{r47}, \cite{pls57}. These high-priority control signals desire reliable and faster transmission among communicating nodes. Therefore, URLLC provides the desired service to these IoT applications by transmitting finite blocklength signals. However, wireless channels are vulnerable to security threats, which can seriously damage the URLLC signal transmission by intercepting or distorting private information.

Generally, cryptography-based algorithms provide effective and secure communication for traditional wireless networks. However, deploying numerous machine-type devices in IoT networks requires lightweight, low-power consuming security solutions with reduced channel overhead \cite{pls98}. Hence, the complex key-based cryptographic security techniques are also unsuitable for supporting URLLC in these IoT applications. As many IoT applications rely on short-packed URLLC signals for the transmission of mission-critical control information, ensuring physical layer security becomes challenging. Additionally, the presence of a large number of low-power IoT devices raises the issue of preserving the overall system energy efficiency of the system along with secure communication and \cite{pls52}, \cite{r47}.

 However, there is a lack of research on providing a lightweight PLS scheme while preserving system energy efficiency in IoT. Only a few works like the ones in \cite{r2} have studied the design of energy-efficient secure short-packet communication in an IoT scenario. The theoretical analysis presented by the work in \cite{r2} only focuses on improving the power allocation scaled to the number of IoT devices. On the contrary, the work fails to consider the intercept probability of the eavesdroppers while ignoring the secrecy constraint to provide secure communication for the FBL URLLC signals. Therefore, from the physical layer perspective, the design of energy-efficient secure transmission schemes for URLLC signals in IoT is still an open research issue.
 
 However, improving spectral efficiency \cite{r3} and energy efficiency of IoT simultaneously is important while proposing PLS schemes for secure data. Considering this, the work in \cite{r3} proposed two strategies, including packet replication and interface diversity, to improve PLS by maximizing the secure spectral efficiency of the short packet URLLC signals in a Nakagami-M fading channel. The work in \cite{r6} has proposed a resource allocation optimization technique for secure communication in mission-critical IoT scenarios. The work aims to maximize the weighted throughput and minimize the transmit power of the system under the secrecy constraint by jointly optimizing the transmit power and bandwidth allocation. Similarly, the performance evaluation of secure URLLC has been studied in \cite{r56} for both single-antenna and multi-antenna cases in an IoT system.

\begin{table}[]
\caption{Required KPIs for URLLC use cases}
\label{tab4}
\begin{tabular}{|l|c|c|c|l|}
\hline

\multicolumn{1}{|c|}{\textbf{\begin{tabular}[c]{@{}c@{}}URLLC\\ Use case\end{tabular}}} & \textbf{\begin{tabular}[c]{@{}c@{}}Reliability\\ (\%)\end{tabular}} & \textbf{\begin{tabular}[c]{@{}c@{}}Latency\\ (ms)\end{tabular}} & \textbf{\begin{tabular}[c]{@{}c@{}}Communication\\ Range (m)\end{tabular}} & \multicolumn{1}{c|}{\textbf{Ref.}} \\ \hline

\begin{tabular}[c]{@{}l@{}}Industrial\\ Automation\end{tabular} & $99.9999999$ & 0.25-10 & 50-100 & \begin{tabular}[c]{@{}l@{}}\cite{pls27},\\ \cite{pls57} \end{tabular} \\ \hline

\begin{tabular}[c]{@{}l@{}}Motion \\Control \\operation \end{tabular}& $99.999999$ & 10-100 & 300-1000 & \cite{r100} \\ \hline

ITS & $99.999$ & 10-100 & 300-1000 & \cite{pls29} \\ \hline

V2X & $99.999$ & 5 & 300 & \cite{pls84} \\ \hline

\begin{tabular}[c]{@{}l@{}}Self\\ Driving\end{tabular} & $99.9$ & $\leq 1$ & 400 & \cite{r72} \\ \hline

Smart Grid & $99.9999$ & 1-20 & 10-1000 & \begin{tabular}[c]{@{}l@{}} \cite{pls55},\\ \cite{pls24}\end{tabular} \\ \hline

\end{tabular}
\vspace{-4mm}
\end{table}

\subsection{Summary and Insights}
 This section discussed the mission-critical use cases, their requirements, and constraints for providing URLLC service. We can understand the importance of URLLC service in providing seamless connectivity. Therefore, considering the security attacks, designing efficient PLS schemes is highly desired while managing the trade-off between security and reliability in all these use cases. 
 
 In summary, we can observe that there exist significant research gaps in providing PLS to URLLC under real-time scenarios of these use cases. Particularly, the physical characteristics of the wireless channel model for these use cases are quite different from each other. So, the PLS techniques like the addition of AN, setting secrecy zones, and using efficient anomaly detection techniques can improve the security of URLLC in these application scenarios. In some cases, the PLS schemes combined with lightweight key-based security enhancement techniques can also be beneficial for URLLC-aided applications \cite{r121}. Furthermore, a detailed characterization of the real-time wireless environment, including a precise identification of attack scenarios for industrial environments, V2X communication, and all such mission-critical applications, have to be considered properly while proposing PLS schemes for URLLC users in their use cases.

\section{PLS of URLLC in 5G Enabling Technologies}

URLLC certainly requires significant enhancements in its security architecture to meet the service requirements of emerging applications. From the physical layer perspective, building lightweight and efficient security enhancement schemes is the key concern of the URLLC system. Apart from this, meeting challenging reliability and latency criteria, multi-user and multi-antenna scenarios, dynamic wireless environment and security vulnerability amplify the complexity of the PLS scheme for URLLC. Therefore, offering data security through enhanced PLS techniques in cutting-edge URLLC enabling technologies like NOMA, MIMO, cooperating communication, and relay-aided (i.e., UAV and IRS) communication has been receiving a noteworthy research interest recently. For better understanding, we review the related studies and comprehensively discuss the potential PLS schemes utilized in these aforementioned URLLC-enabling technologies in the following sub-sections.

\subsection{MIMO and massive MIMO enabled URLLC}

\begin{figure*}[t]
	\centering
	\includegraphics[width=\linewidth, height=5.5 cm]{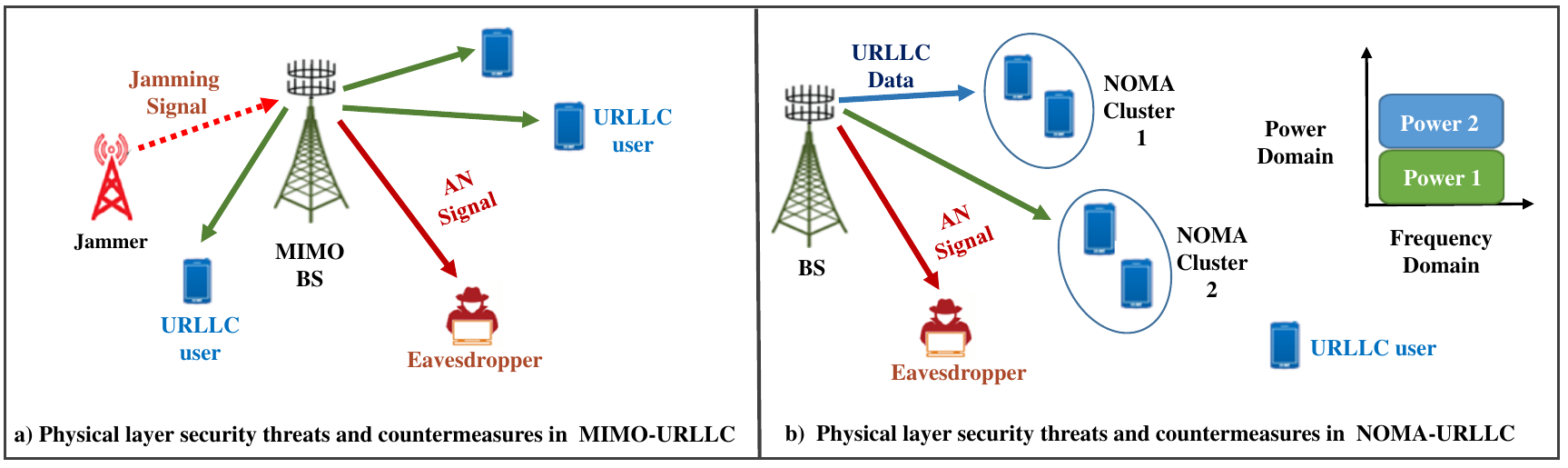}
	\caption{PLS in MIMO and NOMA-aided URLLC.}
	\label{fig6}
	\vspace{-4mm}
\end{figure*}

With the advent of new applications, the next generation of wireless networks is expected to satisfy the requirement of high capacity to accommodate the growing network traffic. In return, this will increase communication complexity and make signal transmission more vulnerable to security threats. Thus, combining PLS techniques and MIMO can effectively provide security to URLLC signals in highly dynamic mission-critical application scenarios. The multi-antenna arrangement in MIMO improves communication reliability by exploiting spatial resources while combating the effect of path loss and interference \cite{r7}. Basically, PLS enhancement in MIMO can be done utilizing a specialized beamforming technique to transmit information-carrying signals in the null space of the eavesdropper channel \cite{r10}. 

To enhance the physical layer security, many investigations have used these multi-antenna technologies for URLLC. Mainly the specialized beamforming technique is utilized to tune the beam toward the intended receiver rather than the eavesdropper for improving the PLS performance. In \cite{r22} the authors have proposed a scenario with a multiple-input and single-output (MISO) system where the transmitter sends the confidential message multiplexed with AN signal to ensure PLS of URLLC in the presence of non-colluding passive eavesdroppers. 

In a multi-user (MU) scenario, MIMO improves wireless channel capacity.  In order to evaluate the performance of MU-MIMO systems under maximum wiretapping by multi-antenna eavesdroppers, an analytical expression for secrecy throughput of finite blocklength signals is presented in \cite{r15}. The analysis considers the self-interference in the full duplex communication scenario of URLLC. Similarly, the impact of fading on the PLS of URLLC in a single-antenna and the multi-antenna case has been studied in \cite{r10}. The work proposes an optimal secure transmission policy by jointly optimizing the finite blocklength and code rate of URLLC to maximize the secrecy throughput. 

Furthermore, the addition of the large antenna array to the MIMO gives rise to a promising new technology called massive MIMO (mMIMO) \cite{r159}. Due to the multi-antenna setting, mMIMO provides several degrees of freedom to wireless networks for improving the quality of signal transmission. Moreover, mMIMO significantly increases the channel capacity to accommodate the growing network traffic and service demand \cite{pls82}. With the space diversity of MIMO and mMIMO, the URLLC systems can achieve maximum diversity gain. Therefore, it is widely adopted and preferred to design the URLLC systems.

In the mMIMO-URLLC system, accurate characterization of the wireless channel is essential for developing an efficient PLS scheme. In \cite{pls81}, the authors provided a framework to characterize the achievable error probability for the mMIMO-URLLC system. The work justified the decrease of error probability to zero when the number of antennas in an mMIMO system increases to infinite with a minimum mean square error processing. The proposed framework considered pilot contamination attack, imperfect CSI, and spatial correlation of channels to provide a non-asymptotic characterization of error probability for both uplink and downlink scenarios of URLLC. The work in \cite{r58} proposed a novel scheme to characterize the shadow fading channel (i.e., $\alpha-k-\mu$ channel) while analyzing the security of the URLLC in terms of average information leakage in the massive MIMO system. The results show that the average information leakage increases as the need for security increases for URLLC.

\subsection{NOMA-aided URLLC}

NOMA is one of the promising enabling technologies for 5G and future 6G wireless networks. It offers improved coverage, high spectral efficiency, and low-latency communication. NOMA allocates radio resources non-orthogonally among users to support massive connectivity in a large-scale wireless network \cite{pls44}. Therefore, low-latency signal transmission is possible in NOMA, making it suitable for URLLC. Due to its high spectral efficiency, it performs better than OMA for short-packet communication. Most importantly, NOMA benefits URLLC by compensating for the performance degradation due to the non-zero decoding error. By exploiting the power allocation, NOMA facilitates reliable communication for cell edge users experiencing bad channel conditions \cite{r42}. 

However, NOMA brings some challenges for URLLC, which cannot be ignored. The major challenge for NOMA URLLC users is to eliminate interference and provide secure data transmission  \cite{pls37}. Basically, NOMA introduces interference in the transmitted signal and eliminates it using successive interference cancellation (SIC) \cite{r113}. In NOMA, the user present at a far distance from the base station (BS) is termed a far user, and those near the BS are called near users. The basic NOMA principle allows near users to decode or demodulate the data of far users in order to apply SIC. This increases the chance of eavesdropping. Therefore, to ensure secure communication, information leakage from near users has to be avoided. At the same time, secure communication is possible by using special coding techniques for the far user data which ensures its security whereas near users can only apply SIC without being able to decode the far user data.  

In NOMA-URLLC, BS can achieve PLS by optimizing the power allocation to the users and also by using an adaptive beamforming technique. This can also address the security and reliability trade-off while providing PLS for URLLC. The communication can take place in both full-duplex (FD) or half-duplex (HD) modes. For URLLC, FD communication is preferred due to its higher spectral efficiency in comparison to HD \cite{pls68}. Apart from this, through multicast transmission, NOMA allocates power to the clusters of URLLC users, whereas in unicast mode, a single user is supported through power allocation. Therefore, NOMA can support multiple URLLC users simultaneously with a fair resource allocation. However, security threats from an internal eavesdropper in the NOMA cluster can launch a jamming attack on the legitimate receiver. As a result, secure data transmission of the URLLC users in the cluster will be hampered, leading to the wastage of allocated power. Hence, these challenges need to be addressed properly while developing PLS schemes for URLLC.

Recently, the research work for enabling URLLC using NOMA has been gaining attention. Prior to this, the research on PLS enhancement through NOMA using energy harvesting jammers \cite{pls37}, hybrid SIC for cognitive radio-inspired NOMA \cite{pls40}, secure relay selection \cite{pls48}, utilizing AN \cite{r112} and cooperative jamming \cite{pls44}, \cite{pls47} have been the focus area for researchers. Based on state-of-the-art research studies, PLS in NOMA has demonstrated its potential in successfully providing secure URLLC signal transmission by accommodating a large number of users simultaneously. In \cite{pls49}, the authors derived the average secure BER for the near user while considering the cell edge user as an untrusted node in a NOMA-assisted URLLC system. The results show that the average secure BER is almost insensitive to the power allocation parameters for both cell-edge and central URLLC users when the average secure BER is less than one. 

However, in case of imperfect CSI, the power allocation in NOMA will be impacted, and the chances of SIC error will increase, which can lead to the degradation of overall secrecy performance. Thus, mitigation of the impact of CSI estimation error on short packet communication needs to be taken into account for developing an efficient PLS scheme. Further, the survey given in in \cite{r113} can be referred for a more detailed and insightful analysis of NOMA-assisted secure URLLC.

\subsection{Cooperative Communication aided URLLC}

Communication through the wireless medium often gets hampered due to the presence of obstacles which can easily block or reflect the transmitted signal. In such a scenario, cooperative communication offers communication with the help of relay nodes to improve the reliability of the wireless signal transmission. Apart from this, the distributed relay nodes in cooperative communication provide the spatial degree of freedom while benefiting the URLLC users in terms of enhanced reliability, energy efficiency, and spectral efficiency \cite{r7}. However, communication through a cooperative relaying scheme needs to address the adverse effects of non-negligible error probabilities in URLLC scenarios. Generally, the relay nodes in cooperative communication use decode-and-forward (DF) or amplify-and-forward (AF) relaying protocol for signal transmission \cite{r20}.  However, the adaptive relaying protocol based on dynamic relay selection (i.e., DF or AF) is useful to satisfy the strict QoS requirements of URLLC \cite{r63}.


Cooperative communication has been a focused area of research for improving the security of short-packed URLLC signals. As we know, the URLLC signals are bound by strict QoS constraints; thus, cooperative communication in such a scenario becomes very effective in achieving the desired QoS level and reducing the signal outage significantly \cite{r28}. Therefore, the proper selection of relay nodes for URLLC signal transmission can improve the overall energy efficiency and provide low-latency communication as well. Thinking about wireless security, these relay nodes in cooperative communication can be utilized as potential jammers for sending AN signals to degrade the quality of the eavesdropping channel \cite{r110}. However, these relay nodes can act as potential eavesdroppers when they become untrusted, making URLLC signals vulnerable to security threats. Therefore, this requires a careful selection of relay nodes to avoid information leakage in high-security conditions.

Several recent studies discuss the advantages of cooperative communication for secure URLLC signal transmission. In some of the research contributions like \cite{r20} and \cite{r28}, short packet secure communication has been investigated in relay-aided wireless networks. The authors of the aforementioned work have utilized cooperative communication strategies to enhance the security and reliability of the URLLC signal transmission under static and mobile conditions of the user. For example, the work in \cite{r28} discusses the role of cooperative communication in providing secure URLLC considering Random Waypoint moving receivers. The work proposes the average bit-error rate (BER) based secrecy evaluation metrics for URLLC while balancing the trade-off between reliability, latency, and security for maximizing the network throughput.

Recently, some of the research works have used aerial relay nodes to transmit the jamming signal to guarantee the reliability and security of confidential information transmission for URLLC. In line with this idea, the authors of the work in \cite{r114} used a friendly jamming technique to ensure PLS for URLLC signal by injecting jamming signals deliberately to the eavesdropping channel through a friendly UAV relay node. The flexible deployment of UAVs here provided a line-of-sight (LOS) channel condition while offering a high-quality air-to-ground channel for improving the PLS of URLLC \cite{uav5}. 


\begin{figure}[t]
	\centerline {\includegraphics[width=1\linewidth, height=4.8 cm]{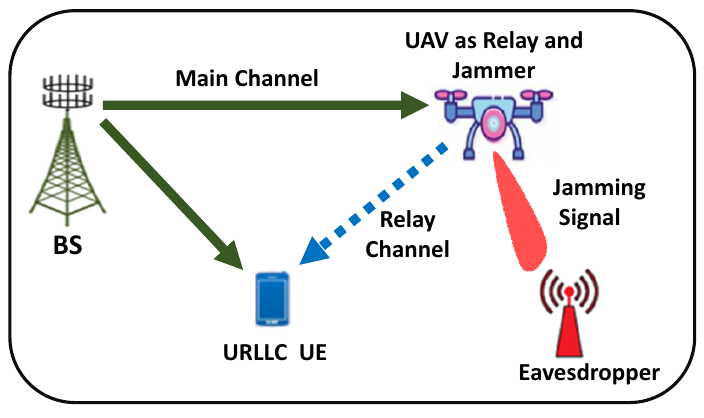}}
	\caption{PLS of URLLC using UAV as relay and jammer.}
	\label{fig7}
	\vspace{-4 mm}
\end{figure}

\subsection{UAV-aided URLLC}

Unmanned aerial vehicles (UAVs) are an integral part of wireless networks. The deployment of UAVs makes the wireless system design flexible and adaptive \cite{pls8}. The on-demand deployment of UAVs establishes reliable and cost-efficient communication in emergency conditions. URLLC signal transmission is hugely benefited by the deployment of UAV relays in the system. UAVs can be used as aerial base stations or can act as relays while providing a line-of-sight (LOS) link for signal transmission. The flexibility offered by UAV becomes helpful for data collection from transmitting nodes in an IoT scenario, thus providing connectivity in extreme environments \cite{uav2}. Specifically, for 6G URLLC applications, handling huge data collected from remote devices and ensuring security simultaneously will become a challenging task.

Generally, UAV-aided communication is affected by two types of eavesdropping attacks such as passive or active eavesdropping \cite{uav2}. In passive eavesdropping, the eavesdroppers intercept the confidential information exchanged between the legitimate transmitter and receiver without degrading the received signal quality at the intended receiver. On the other hand, active eavesdroppers deliberately interfere with the confidential signal transmitted by UAVs and degrade its quality by deliberate jamming. Therefore, active eavesdropping is more harmful to UAV-aided URLLC signal transmission in comparison to passive eavesdropping \cite{r101}.

Despite all the advantages offered by UAVs in ensuring reliable, secure, and low-latency communication, we cannot deny the adversaries of UAV-aided communication. The UAVs can increase the chances of interference among aerial and ground user equipment which can further degrade its performance \cite{uav3}. Additionally, UAVs are energy-constrained devices. Therefore, energy harvesting becomes essential while using UAVs in cellular infrastructure. UAVs can also act as jammers to protect the communication between legitimate transmitters and receivers \cite{r122}. Recently, the PLS in UAV-assisted URLLC has been studied in \cite{uav5} for a massive MIMO system. The work utilizes a path loss model for air-to-ground links between UAV and ground users· A secrecy guard zone has been proposed for secure communication in the UAV-enabled network. It has been observed that secure communication improves with the increase in the radius of the secrecy protection zone, while it has been negatively affected by the increase in eavesdropper density. However, PLS improvement in UAV-aided URLLC has to face several challenges, like high mobility, dynamic channel conditions, and NLOS scenarios \cite{irs12}, \cite{uav6}. 

Moreover, UAVs are energy-constrained devices, and their deployment in the wireless network requires extra power allocation. Therefore, the PLS of UAV-aided URLLC systems require lightweight PLS solutions to make the communication process energy efficient and secure simultaneously. Therefore, efficient energy harvesting schemes are needed to support the operations of UAV-aided communication \cite{pls74}. This requires proper user scheduling and URLLC data transmission process via the joint optimization of parameters like blocklength, transmit power, and UAV trajectory under secrecy constraints to develope a feasible PLS solution for secure UAV-aided communication \cite{uav5}. 

Recently, research studies related to the use of UAVs in providing URLLC service have been on the rise. In this regard, a comprehensive survey on UAV-enabled URLLC is present in \cite{uav7}. The work highlights the requirement of UAV-enabled communication to satisfy the QoS requirement of URLLC while emphasizing the requirement of secure communication through PLS schemes. Short packet communication from the sensors to a remote ground BS with the help of UAVs has been studied in \cite{uav2}. The authors have a secure URLLC signal transmission framework while preserving the freshness of data collection and transmission via UAVs. The work suggests a joint optimization framework of user scheduling, UAV trajectory, and flight duration optimization to maximize the energy efficiency of UAV-aided communication. Furthermore, in the data transmission phase of UAV, the maximization of secrecy rate has been achieved by jointly optimizing transmit power and blocklength under the constraints of eavesdropping rate and outage probability of UAV to BS link. Similarly, in \cite{uav1}, the UAV acts as an aerial relay node that decodes and forwards the URLLC control signals from the source node to the actuator devices in the presence of external EVs. The authors have adopted NOMA-based SPC to improve coverage, reduce latency, and achieve fairness among users. The work proposes a joint optimization URLLC signal block length, transmit power, and UAV trajectory to maximize the minimum average secrecy throughput of all actuator devices using BCD and successive convex approximation methods.

\subsection{IRS-assisted URLLC}

IRS-aided communication is considered as an emerging technology to support URLLC in 5G and upcoming 6G \cite{r117}. Due to the high energy efficiency and enhanced coverage extension capability, it becomes suitable for enabling URLLC applications in non-line-of-sight (NLOS) communication scenarios \cite{r158}. Furthermore, IRS is capable of redirecting the signals towards the receiver intelligently with the help of a large number of passive reflective elements. IRS-enabled communication is energy efficient because it can steer the phase of the incident signal towards the receiver without utilizing any energy \cite{irs3}. Thus, it is perfectly suitable for large-scale IoT applications like industrial IoT that rely on transmitting short packet URLLC signals among a large number of low-power devices \cite{irs4}. 

More importantly, IRS is capable of configuring the wireless propagation environment by controlling the direction and gain of the beam towards the desired receiver \cite{irs14}. This inherent characteristic of IRS helps in improving the PLS of the URLLC signal transmission by constructively adding user signals for the legitimate URLLC receivers while destructively adding signals for the eavesdroppers \cite{irs1}. In recent years, research studies related to the use of the IRS for providing URLLC service have been on the rise. The studies like \cite{irs1}, \cite{irs3}, and \cite{irs5} have proposed efficient techniques utilizing IRS to support URLLC while improving the energy efficiency of the system. For example, the survey in \cite{irs15} focuses on the key security challenges for IRS-aided communication from the perspective of future 6G applications. The authors discussed various scenarios of IRS-assisted communication in millimeter wave and terahertz communication, large-scale IoT networks, edge computing, and non-terrestrial networks that are vulnerable to security threats and jamming attacks. The survey studied and discussed various types of attacks or security threats that can hamper IRS-aided communication. 

Additionally, the work in \cite{irs2}, discussed the security of IRS-aided communication scenarios while identifying some key future research directions to counter potential security threats. Similarly, the survey in \cite{irs14} also provided a comprehensive survey of IRS-empowered future 6G wireless communication while discussing various IRS deployment scenarios like centralized and/or decentralized strategies in 6G. Moreover, the optimization techniques used in wireless technologies like MIMO, NOMA, and multi-access edge computing with IRS are integrated and discussed in the study. However, the discussion on privacy and security of communication through the IRS is missing in the work. Apart from this, the use of IRS for the systems using URLLC and the PLS techniques addressing the possible security threats in such mission-critical scenarios were not explored in the survey. 

There are very few research works have been carried out to study the security enhancement of URLLC in IRS-aided communication. For example, the work in \cite{irs12} has discussed various use cases of IRS-aided UAV communication where PLS can be ensured using the IRS as a relay.  A case study has been provided in the work to maximize the secrecy capacity of the UAV-IRS integrated network in the presence of external eavesdropping. In another work presented in \cite{r117}, the authors have proposed a self-secure finite blocklength coding scheme for IRS-assisted communication in the presence of multiple eavesdroppers. This scheme considers the availability of NLOS channel conditions between the legitimate transmitter and receiver in different antenna settings like SISO, SIMO, and MIMO systems. The work emphasizes the requirement of accurate CSI estimation, which is essential for efficient beam forming to enhance secure communication through the IRS. Sometimes, IRS-aides communication is adversely affected by intentional pilot contamination. As a result, the transmitter inefficiently designs the IRS beamforming with pilot contamination, which ultimately increases the information leakage \cite{r119}. In some scenarios, estimating accurate CSI becomes challenging because of the passive IRS elements which cannot transmit and receive signals actively to estimate the channel parameters \cite{irs16}. Therefore, developing efficient PLS schemes for IRS-assisted URLLC needs to overcome these challenges in 5G and future 6G applications.

\begin{figure}[t]
	\centerline {\includegraphics[width=1\linewidth, height=4.9 cm]{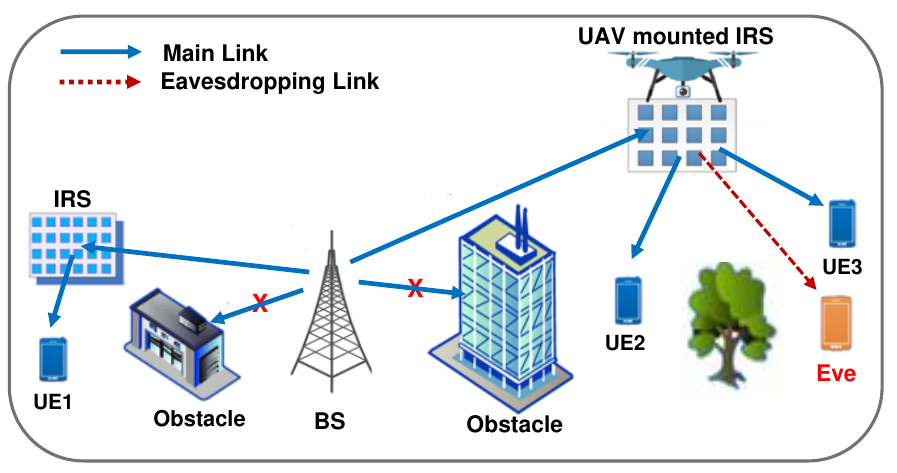}}
	\caption{PLS in UAV and IRS assisted URLLC.}
	\label{fig8}
	\vspace{-4 mm}
\end{figure}

\subsection{IRS integrated UAV-aided URLLC}

Integration of UAV and IRS brings several benefits by improving reliability, connectivity, and coverage of signal transmission. Most importantly, IRS-aided UAV relays can minimize the transmission delay of URLLC signals by eliminating the requirement of two-time slots for decoding and forwarding the signal earlier used for UAV-aided URLLC \cite{pls56}. Moreover, power consumption is minimized significantly due to the presence of the IRS, which acts as a passive relay to forward the URLLC signal. IRS can potentially improve the PLS of UAV-aided URLLC communication by transmitting AN signals toward eavesdroppers and constructively adding the transmitted confidential information for legitimate users \cite{pls6}. Apart from this, the challenge of an accurate user positioning of IRS-aided communication can also be eliminated by using UAVs to successfully map the eavesdropper's position. 

 IRS mounted on UAVs provide directional beamforming towards the legitimate receiver and avoid information leakage. Therefore, an integrated IRS-UAV network can be utilized as a green jammer for eavesdroppers by producing jamming signals without any external energy supply. The use of IRS in UAV-aided communication also improves PLS and maximizes secrecy capacity in the presence of multiple Eves \cite{r119}. Recently, the work on integrating IRS with UAV for facilitating URLLC is gaining attention. The authors in \cite{r61} utilized a UAV-mounted IRS as an aerial passive relay for maximizing the secrecy rate of the communication by jointly optimizing the UAV trajectory, IRS passive beamforming, and power allocation. However, the aforementioned work utilizes long blocklength communication. However, we can infer that the integration of IRS into UAV-aided communication to design energy efficiency PLS scheme for URLLC. This can also benefit URLLC signal transmission by providing flexible deployment in mission-critical application scenarios and simultaneously ensuring the PLS of the control information.

\subsection{PLS of URLLC using Secrecy Guard Zone}

\begin{figure}[t]
	\centerline {\includegraphics[width=0.8\linewidth, height=6 cm]{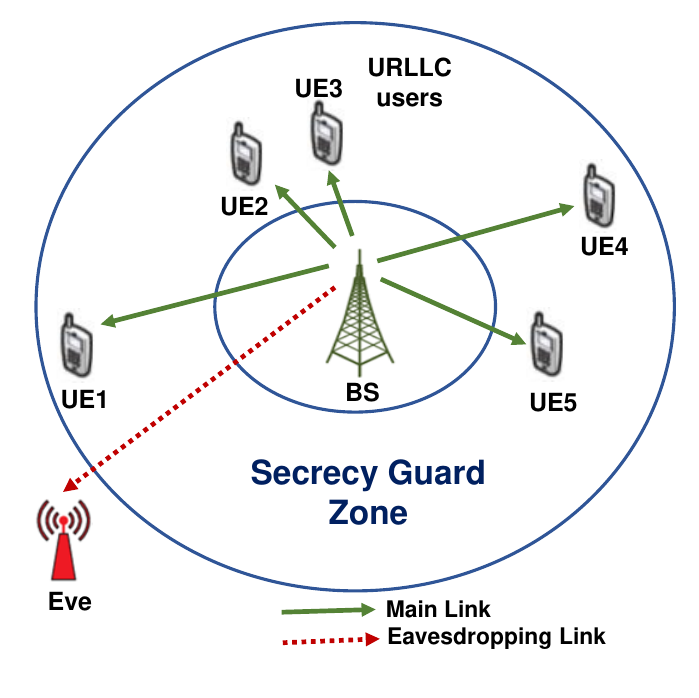}}
	\caption{PLS of URLLC using Secrecy Guard Zone.}
	\label{fig9}
	\vspace{-4 mm}
\end{figure}

The knowledge regarding the physical presence of eavesdroppers and the location with respect to the receiver is another important factor in ensuring secure communication for the wireless system. Physically close eavesdroppers can significantly hamper the ongoing communication and degrade the required reliability of the URLLC signal in a scenario where multiple users are served by a single BS in a cluster-like area. In such conditions, one of the effective techniques to provide secure signal transmission to the intended user is by creating a Secrecy Guard Zone (SGZ) surrounding the transmitter \cite{r34}. Basically, SGZ is an area created around the transmitter without any eavesdroppers and containing only legitimate users. The radius of the SGZ can be optimized according to the location of users and the QoS requirements. 

The secrecy zone is created by varying the transmit power at the transmitter. This technique is also applied to URLLC users looking for secure communication.  In general, SGZ-based PLS enhancement considers two signal transmission policies in which the transmitter can remain silent when the eavesdroppers are present in the SGZ or transmit the AN signal instead \cite{r31}. On the other hand, signal transmission only occurs when eavesdroppers are outside the SGZ. Therefore, SGZ can effectively provide secure URLLC and improve secrecy throughput for mission-critical wireless networks that have high-security requirements. In line with this, the work in \cite{uav5} has proposed an SGZ to safeguard the URLLC signal transmission in the presence of eavesdroppers. Here, the AN signal is transmitted from the multi-antenna transmitter along with an information signal to protect the confidentiality of the URLLC signal. In this condition, optimization of the power allocation factor to both the information signal and AN signal under URLLC constraint is highly desired from the SGZ design perspective.

\begin{table*}[t!]
\caption{Summary of PLS enhancement techniques for URLLC}
\label{tab5}
\begin{center}\begin{tabular}{|l|l|l|}
\hline
\multicolumn{1}{|c|}{\begin{tabular}[c]{@{}c@{}}URLLC \\Enabling Technology\end{tabular}} & Research & Key Points \\ \hline

\multirow{2}{*}{ MIMO and mMIMO} & \cite{r22}  &  AN used to counter passive eavesdropping in MISO-URLLC system.\\ [1mm] \cline{2-3}
& \cite{r10} & Specialized beamforming used to transmit information in the null space of the eavesdropper channel \\ [1mm] \cline{2-3} 
& \cite{r15}  & PLS enhancement for multi-user MIMO avoiding self-interference. \\ [1mm] \cline{2-3}
& \cite{pls83}  &  Artificial noise signal used to improve the URLLC system average secrecy throughput. \\ [1mm] \cline{2-3}
& \cite{r58}  & Improved PLS of URLLC in $\alpha-k-\mu$ shadowed fading channel reducing average information leakage \\ [1mm] \hline

\multirow{3}{*}{NOMA } & \cite{r111} & Secure communication achieved using special coding techniques for URLLC users present far from BS.\\ [1mm] \cline{2-3}
& \cite{r112} & Achieve PLS by optimizing the power allocation and adaptive beamforming. \\ [1mm] \cline{2-3} 
 & \cite{pls68} & Provides a comprehensive survey of the PLS-assisted NOMA and the state-of-the-art related research works. \\ [1mm]  \hline

\multirow{3}{*}{Cooperative communication} & \cite{r20} & PLS enhancement through cooperative relay considering user mobility. \\ [1mm] \cline{2-3} 
& \cite{r28} & Average BER-based secrecy evaluation for Random Waypoint moving URLLC users. \\ [1mm] \cline{2-3}
& \cite{r114} & Used friendly cooperative relaying scheme for transmitting jamming signal to eavesdroppers. \\ [1mm] \hline

\multirow{2}{*}{UAV-aided URLLC} & \cite{pls50} & Facilitate secure URLLC, improve secrecy throughput using block successive convex approximation. \\ \cline{2-3}
& \cite{uav5} &   Minimize secrecy outage probability and connection outage probability. \\ [1mm] \cline{2-3} 
& \cite{uav1} &  Maximize minimum average secrecy throughput.  \\ [1mm] \cline{2-3} 
&  \cite{uav2}& Maximize EE, guarantee information freshness for secure URLLC.  \\ [1mm] \cline{2-3}
&  \cite{uav6}  & Improve secrecy rate for URLLC.  \\ [1mm] \cline{2-3}
& \cite{uav8} & Joint optimization of URLLC blocklength, transmit power, and UAV’s trajectory to achieve PLS. \\ [1mm] \cline{2-3}
& \cite{r122} & UAVs used as a friendly jammer to achieve PLS. \\ [1mm] \cline{2-3}
& \cite{uav4}  & Weighted secrecy coverage enhancement using UAV as a friendly jammer. \\ [1mm] \hline

\multirow{2}{*}{IRS-aided URLLC} & \cite{r113} & Improve PLS for an IRS-aided NOMA with and without eavesdropper CSI. \\ [1mm] \cline{2-3}
& \cite{r117} & Self-secure finite coding scheme used to improve PLS. \\ [1mm] \cline{2-3} 
& \cite{r119}  & CSI-based accurate beamforming used to improve PLS. \\ [1mm] \cline{2-3}
& \cite{r158} & Provide a case-study based analysis of IRS-enabled URLLC applications, challenges, and potentials \\ [1mm] \cline{2-3}
& \cite{irs3} & Support wireless energy transfer and information transfer for URLLC. \\ [1mm] \hline

\multirow{2}{*}{ IRS-UAV integrated URLLC } & \cite{r61} & Improve secrecy rate by jointly optimizing UAV trajectory, IRS phase shift element, and power. \\ [1mm] \cline{2-3} 
& \cite{pls56} & Ensure secure data offloading using UAV-integrated IRS. \\ [1mm] \cline{2-3}
& \cite{irs15} & Survey on secrecy and privacy of IRS-assisted communication in 6G.  \\ [1mm] \hline

\multirow{2}{*}{Secrecy Guard Zone} & \cite{uav5}  &  Improve secrecy rate and minimize outage probability using SGZ in MIMO-URLLC network. \\ [1mm] \cline{2-3} 
& \cite{r31}  &  Improve secrecy capacity using SGZ for long blocklength codes. \\ [1mm] \hline


\end{tabular}
\vspace{-5 mm}
\end{center}
\end{table*}

\subsection{Summary and Insights}

In summary, this section provides a comprehensive study of adopted PLS techniques to counter security attacks and prevent information leakage in various URLLC enabling 5G techniques Like MIMO, mMIMO, NOMA, and relay-aided cooperative communication from a physical layer perspective. Although some significant research efforts have been made to enhance PLS for short packet communication, dynamic and heterogeneous service requirements with conflicting QoS demands remain a great challenge from a security perspective. In the case of MIMO and mMIMO, growing data traffic, uncertainty in CSI of eavesdroppers, PCA attacks, and jamming attacks will introduce challenges and demand for lightweight PLS schemes. 

Relaying has been proven as an efficient technique to provide reliable and low-latency communication while simultaneously preserving data security by acting as a friendly jammer and transmitting AN signals. Therefore, exploring the new possibilities through UAV and IRS-aided secure URLLC signal transmission can enhance conventional PLS mechanisms to adapt to a more dynamic and heterogeneous wireless environment. However, the complexity in parameter optimization (i.e., UAV trajectory, IRS phase shift matrix, and transmit power) in cooperative relaying through UAV and IRS is required to be handled efficiently in multi-user and multi-eavesdropper scenarios. with the transformation of URLLC enabling technologies, security issues will become more prevalent. As a result, dynamic security enhancement policies, transmission protocols, and security countermeasures need to be developed to elevate the performance of PLS techniques.

\section{ML enabled PLS solutions for URLLC}

Facilitating URLLC is crucial for many mission-critical applications. In this regard, the inclusion of ML-based techniques provides potential benefits to achieve the desired QoS levels for these applications \cite{r137}. URLLC networks can be potentially benefited by leveraging the facilities offered by ML to enhance the system performance. This allows for the provision of ML-enhanced URLLC by potentially optimizing resource allocation based on real-time wireless network condition and QoS demands in mission-critical services and applications across several domains. For ensuring URLLC security, ML algorithms can be utilized efficiently to detect anomalies by analyzing historical data patterns and real-time monitoring of the wireless network. However, the challenges incurred due to the complex wireless environment, heterogeneous service requirements, and exponentially increasing network traffic introduce roadblocks to the implementation of ML-based solutions to provide URLLC in these applications. Hence, designing dynamic and intelligent ML-based PLS schemes need to address the aforementioned challenges to facilitate secure URLLC service.
 
Generally, the centralized ML model that facilitates and controls URLLC signal transmission needs to use pre-trained ML models or an online training process that can significantly reduce the end-to-end delay \cite{r141}. In such conditions, preserving the security and privacy of mission-critical URLLC signals becomes very much important as they can be easily hampered by cyber attacks. In this regard, survey works like \cite{pls62} and \cite{r120} provided a comprehensive review of the ML and deep learning (DL) enhanced PLS techniques for establishing secure communication in 5G and beyond wireless networks. Specific to URLLC, recent work in \cite{r70} has proposed a security solution for the physical layer and higher communication protocol layers in case of short packet URLLC signal transmission. The proposed approach was flexible enough to satisfy URLLC QoS and security requirements simultaneously. The authors of the work used stochastic network calculus to model the secure transmission scheme for URLLC. Then, the formulated optimization problem was solved by using an experienced Meta Asynchronous advantage actor-critic (EM-A3C) algorithm, which is capable of establishing an experience pool for improving the efficiency of the proposed algorithm.

Moreover, the coexistence of multiple 5G services like eMBB and mMTC with URLLC in mission-critical applications introduces challenges in developing efficient PLS techniques due to their diverse QoS requirements like high data rate, high reliability, large connection density, and low latency communication simultaneously. In such a scenario, security threats like “denial of service" can introduce challenges in satisfying service requirements and PLS design simultaneously for URLLC \cite{r73}. Furthermore, estimating accurate CSI in such heterogeneous networks makes it difficult to establish secure communication among end users. Therefore, the knowledge of statistical CSI is often utilized for designing ML-enabled intelligent power control strategies for mitigating several eavesdropping attacks in these heterogeneous networks with multi-service coexistence of URLLC\cite{irs12}.

Recently, ML has been employed successfully by researchers to achieve secure data transmission and enhance the capability of conventional PLS schemes. The use of ML techniques such as DRL for optimizing beamforming \cite{pls88}, \cite{pls89}, beamforming using DNN for improving secrecy rate in the UAV-aided communication \cite{pls90}, optimal relay selection using decision tree \cite{pls91} and RL-based relay selection and power allocation \cite{pls92} have contributed in enhancing the overall PLS performance for long block length codes. Therefore, with optimal parameter selection and ML model tuning, these ML-based PLS enhancement tuning, techniques will be able to achieve desired security levels for finite blocklength URLLC.

Basically, ML models require large training datasets to provide dynamic and intelligent solutions for solving wireless communication challenges. In most cases, training of the ML model is a centralized process and requires high computational capability for processing such large amounts of data \cite{r139}. However, the centralized ML models present in the centralized servers are prone to hacking, as breaking into the single central server is easier. This increases the chance of information leakage and alteration of control information in URLLC-enabled critical applications. Thus, manipulating the control information through proactive hacking can have serious consequences in terms of increased road accidents in ITS or manufacturing defects in industrial process control. Therefore, careful model deployment and the use of intelligent security protocols for critical information processing for URLLC applications will become the focus point of many researchers.

\begin{table*}[t!]
\caption{Summary of ML-enabled Security Enhancement Techniques for URLLC. }
	\label{tab6}
\begin{center}
\begin{tabular}{|l|l|l|l|}
\hline

\multicolumn{1}{|c|}{\textbf{Research}} & \multicolumn{1}{c|}{\textbf{Objective}}  & \multicolumn{1}{c|}{\textbf{ML Model}}  & \multicolumn{1}{c|}{\textbf{Key Features}}  \\ \hline

\cite{pls35} & Secure 5G   &  Federated DRL   & \begin{tabular}[c]{@{}l@{}} Mitigating jamming attack in highly dynamic 5G heterogeneous network.\end{tabular}   \\ \hline

\cite{r72} & Secure URLLC   &  Federated RL  &  \begin{tabular}[c]{@{}l@{}} Task offloading by assessing RSU’s reputation score using global network obtained from \\A3C algorithm in a three-layer architecture (edge server, convergence server, and \\cloud server).\end{tabular} \\ \hline

\cite{irs12} & Secure 5G & Q-Learning    & \begin{tabular}[c]{@{}l@{}} The interactions among legitimate transmitter and eavesdropper modeled as a zero-sum \\game for eavesdropping, jamming, and spoofing attacks with \\statistical CSI knowledge of the attacker.\end{tabular}   \\ \hline

\cite{r70} & Secure URLLC  & \begin{tabular}[c]{@{}l@{}} Experienced Meta \\Asynchronous advantage \\actor-critic (EM-A3C)\end{tabular} & \begin{tabular}[c]{@{}l@{}} Cross-Layer Security Solution with initiative waiting strategy using flexible TTI-based \\scheduling and flexible pre-backup transmission.\end{tabular}  \\ \hline

\cite{pls5} & Secure 5G   & Federated Learning   & \begin{tabular}[c]{@{}l@{}}Mitigate the security threats due to poisoning and membership inferences attacks in a \\decentralized 5G network\end{tabular}   \\ \hline

\cite{pls53} & Secure 5G IoT   & Deep Learning   & \begin{tabular}[c]{@{}l@{}} Improve trust and confidence in anomaly-based intrusion detection in 5G IoT networks\\ using explainable AI technique.\end{tabular}   \\ \hline

\cite{pls33} & Secure URLLC   &  Model based DRL   & \begin{tabular}[c]{@{}l@{}} Intelligent multi-tier model-based DRL framework proposed to minimize queuing and \\transmission delay in an URLLC aided V2X communication.\end{tabular}   \\ \hline

\cite{r64} & Secure URLLC & Federated DRL  & \begin{tabular}[c]{@{}l@{}} Introduced privacy leakage degree and action relation for improving detection accuracy \\of Anomaly detection model in IIoT.\end{tabular}  \\ \hline

\end{tabular}
\vspace{-5 mm}
 \end{center}
\end{table*}

\subsubsection{Distributed ML architecture for Secure URLLC}

For wireless communication, distributed and decentralized ML architecture significantly reduces the latency of operation and signal transmission delay. Basically, the centralized processing of user data is vulnerable to security attacks because of a single point of operation and incurs serious threats to personal information utilized for training purposes \cite{r140}, \cite{pls64}. To counter this, decentralization of the ML models provides guaranteed security against hacking and eavesdropping of high-priority URLLC control information in large-scale wireless networks \cite{pls102}. In this regard, federated learning (FL) models are utilized successfully to establish secure communication between the central server and the distributed local servers where the model updates are only transferred instead of the personal data of local end-users \cite{r126}, \cite{pls65}, \cite{pls51}. The use of FL in facilitating URLLC and preserving the privacy of the data has been explored by many researchers.

Considering the growing demand for URLLC service in various applications, the development of secure ML models to facilitate URLLC becomes significant. As a promising use case, industrial IoT (IIoT) revolutionized the current industrial landscape by improving production efficiency and reducing the cost of manufacturing with the inclusion of AI-based technologies and 5G connectivity. However, the wide distribution of IoT devices in IIoT becomes vulnerable to security threats if proper protection techniques are not adopted \cite{r57}. In this regard, the development of an efficient anomaly detection or intrusion detection strategy \cite{pls77} with an effective attack anticipation scheme will be helpful for establishing secure communication for control information transmission among URLLC IoT devices present remotely to counter security attacks \cite{r125}, \cite{pls32}. One such method is proposed in \cite{r64}, where an FL-based technique is utilized to develop an anomaly detection global model for the large-scale IIoT system. In this proposed privacy-preserving anomaly detection model, the local models are trained using the deep RL (DRL) algorithm. The work achieves the desired QoS requirements and enhanced accuracy in detecting anomalies for privacy protection in critical URLLC information exchange in the IIoT scenario.

In 6G, vehicle-to-everything (V2X) communication will provide massive connectivity on the road to the roadside units, other vehicles, edge access points, and cellular base stations \cite{pls33}, \cite{pls34}. Specifically, these V2X nodes use URLLC signals to transmit mission-critical control information to each other. The control information exchanged in V2X is mostly processed through onboard AI-enabled devices. However, control information exchange and vehicular task offloading have to face the problem of zero trust networks \cite{r72}. Basically, the concept of zero trust network is introduced to counter the security threats due to malicious attacks from untrusted edge nodes in V2X communication. Considering the unreliable communication and information leakage in URLLC-aided 6G vehicular communication, the work in \cite{r72} has proposed a privacy-preserving federated reinforcement learning (FRL) algorithm for secure task offloading, resource slicing, and efficient scheduling of URLLC signal. The use of the FRL model for secure task offloading makes the system adaptable to dynamic network wireless environment and security threats. Extending the use of ML-enables security solution, the work in \cite{irs11} has utilized a deep reinforcement learning (DRL) framework for enhancing the PLS  of an IRS-assisted network. Basically, the authors tried to maximize the energy efficiency of the system under the secrecy criteria by jointly optimizing the beamforming vectors and the artificial noise vectors for legitimate receivers and eavesdroppers simultaneously. The proximal policy optimization (PPO) algorithm is utilized in this work to maximize the secure energy efficiency of the IRS-assisted network. More details on the ML-enabled PLS solutions can be seen in the short survey given in \cite{r120}.

\section{Hyper Reliable Low Latency Communication in 6G}

The ambitious vision of 6G is to set the foundation for a truly autonomous network that can expand the horizon of current 5G applications and network capabilities. The 6G wireless system envisions unprecedented technological advancements with the emergence of futuristic applications like tactile internet, autonomous driving, industry 5.0, and internet-of-everything (IoE) \cite{pls60} integrating advanced technologies like holographic teleportation, XR applications, quantum communication, and AI \cite{r136}, \cite{r133}, \cite{r138}. This integrated communication landscape opens the possibility of connecting billions of smart devices and end users wirelessly while supporting the ever-growing network traffic and service demand simultaneously \cite{pls42},\cite{r108}. 

Recently released IMT-2030 recommendation includes the proliferation of advanced cutting-edge technologies in 6G like terahertz (THz) communication \cite{pls58}, extreme ultra-massive MIMO (UM-MIMO), communication through Non-Terrestrial Networks (NTN), and Al integrated network for ubiquitous connectivity and communication \cite{pls101}. To support this, an extension to the IMT-2020 capabilities and 5G services has been recommended including new services such as HRLLC, immersive communication, and massive communication for 6G. From the 6G vision perspective, HRLLC is expected to support the extreme reliability (more than 99.999$\%$ ) and latency (less than 1 ms ) \cite{pls93} criteria while simultaneously facilitating the envisioned extreme key performance indicators (KPIs) compared to 5G \cite{r138}, \cite{r134}, as shown in Table \ref{tab7}.  




6G communication will always be strict on preserving the security of data transmission of the large volume of information generated from billions of interconnecting smart devices \cite{pls58}, \cite{pls41}. Most importantly, in all such scenarios, preserving data confidentiality by enabling ultra-high security along with the individual QoS requirements becomes challenging for 6G networks. Moreover, with the advancements in technology, the complexity of the 6G network is also expected to increase multifold. Therefore, supporting extreme QoS requirements of HRLLC with the help of 6G cutting-edge technologies needs to understand the inherent challenges in providing secure data transmission.  For example, applications enabled by HRLLC depend on short-range communication between access points and nearby robots in industrial automation,  wearable devices to users in smart healthcare systems, and road-side units (RSUs) to autonomous vehicles in ITS using communication sub-networks \cite{pls93}. In such scenarios, the presence of strong eavesdroppers in sub-networks can easily intercept confidential information because of better channel conditions and dense deployments. Thus, relying only upon physical layer parameters for the development of advanced security mechanisms will not be sufficient; hence knowledge regarding one of the essential requirements for HRLLC use cases in 6G. 

Hence,  adaptive and intelligent security enhancement schemes need to be developed by integrating AI-based solutions and should encompass multi-layer communication protocols considering HRLLC requirements in various 6G applications \cite{r132}. In this regard, we explore the possible emerging advanced technologies that can provide enhanced security in terms of the physical layer and cross-layer security requirements for HRLLC by analyzing the complexity of the 6G network and understanding the threat landscape. The following subsections summarize some key potential candidate technologies that can be utilized efficiently to provide security to URLLC in 5G and HRLLC in 6G applications.

\begin{table}[t!]
        \caption{KPIs 5G vs. Future  6G}
		\label{tab7}
		\begin{tabular}{|l|l|l|}
			
			\hline
			\textbf{KPI }   & \textbf{5G} & \textbf{6G} \\ \hline
			\centering
			Peak Data rate: Downlink    & 20 Gbps & 1 Tbps \\ \hline
			\centering
			Peak Data rate: Uplink      & 10 Gbps   &  1 Tbps    \\ \hline
			\centering
            Reliability        & Upto 99.999$ \%$    & Up to 99.99999$ \%$  \\ \hline
			\centering
   		User Plane Latency          & $0.5ms$  &    $0.1-1ms $    \\ \hline
			\centering
			Control Plane Latency       & $10ms$  &    $<1ms $    \\ \hline
			\centering
			DL spectral efficiency    & 30 b/s/Hz  &  100 b/s/Hz      \\ \hline
			\centering
			Connection Density        & Upto $10^{6} $    & Up to $ 10^{8} $  \\ \hline
			\centering
			Mobility         & Upto 500 km/h     &   Up to 1000 km/hr     \\ \hline
			\centering
			Operating frequency     & 3-300 GHz &   Up to 1 THz    \\  \hline
			
		\end{tabular}
		\centering
	\vspace{-6 mm}
	\end{table}

\subsection{Quantum enabled Security in 6G}

To enable secure HRLLC in 6G, the conventional PLS techniques may not be efficient enough due to the network complexity and stringent QoS requirements. In such a mission-critical scenario, quantum communication has emerged as a potential candidate to provide secure communication by developing novel security enhancement techniques. Quantum computing (QC) offers unique advantages over classical computing in terms of certain cryptographic protocols, which can effectively address wireless communication challenges \cite{pls61}. Therefore, it has the potential to enhance the security of URLLC/HRLLC signals in several ways. QC can speed up the process of solving complex optimization problems of communication networks, such as load balancing, scheduling, and resource allocation. This can potentially lead to satisfying the lower latency criteria of URLLC/HRLLC \cite{q1}. At the network level, QC can offer enhanced security features for protecting confidential information against various cyber threats. However, quantum computers can decrypt certain classical encryption algorithms that are used for securing wireless communication by extracting sensitive information. Therefore, efficient quantum learning-based frameworks and algorithms need to be developed to ensure secure mission-critical communication in future 6G applications \cite{q2}.

Basically, URLLC supports the grant-free transmission scheme to satisfy its low latency criteria. In grant-free transmission, BS allocates different pilot resources to the URLLC users for channel estimation and identifies the colliding users that share the same time-frequency resource. In such conditions, pilot signals become easily vulnerable to security attacks. The high risk lies in identifying the uncertainty of these pilot-aware attacks on finite blocklength URLLC signals, which can easily paralyze the wireless grant-free transmission. To address this, the authors in \cite{q2} have proposed a non-random superimposed coding scheme for encoding and decoding the pilot signals in a grant-free uplink URLLC scenario. A quantum learning network is utilized in the work for detecting the user activity and pilot-attack mode. Particularly, the quantum learning network exploited the URLLC signal features to develop a multilayer decoding network and remove the uncertainty in the decoding process in the proposed coding scheme under pilot-aware attacks.

In line with the recent advancements in the field of quantum-enhanced secure URLLC service, the researchers in \cite{q1} have proposed a distributed quantum communication protocol that allows the legitimate transmitter and receiver to apply a controlled unitary operation on any input qubit state without sharing the entanglement and unitary operators previously. To ensure security, the physical transmission of particles over the quantum channel was prohibited. Basically, to ensure the security of the massive amount of private data in a large-scale URLLC network, distributed learning models are leveraged. This helps the processing of user information at local servers rather than using centralized models. This multi-tier computing process helps to ensure reliable and low-latency communication in 6G.

Thinking about HRLLC in 6G, post-quantum cryptography (PQC) is another emerging security enhancement technique that can provide secure communication to counter quantum-enabled security threats in the future \cite{bernstein2017post}.  PQC is a method of designing specific cryptographic algorithms and protocols to resist attacks emerging from quantum computers \cite{q3}, \cite{q6}. Generally, quantum computers have the potential to solve certain mathematical problems, such as factoring large numbers and computing discrete logarithms much more efficiently than classical computers. Utilizing these capabilities, PQC focuses on developing efficient encryption schemes that are resistant to security threats \cite{q5}. Therefore, many researchers recently explored PQC to address the security challenges of wireless communication and provide secure data transmission by exploiting the physical channel parameters in the signal encoding and decoding processes. 

Security of URLLC/HRLLC signal transmission is crucial for many mission-critical IoT applications. However, protecting the initial access of URLLC-enabled massive IoT networks in case of signal tampering by malicious quantum adversaries is still challenging. To address this a nested hash access system is proposed in \cite{q4} using the post-quantum encryption on multi-domain physical-layer resources to avoid passive eavesdropping. The proposed technique was able to achieve a higher privacy level using compression and encryption mechanisms based on a quasi-cyclic moderate-density parity-check method. Therefore, in future 6G URLLC applications, the long-term security of communication channels can be ensured by implementing PQC methods for secure URLLC/HRLLC service.

\subsection{Machine Unlearning for Preserving Security and Privacy in 6G}

Basically, ML models are quite capable enough to identify and mitigate external eavesdropping attacks to make URLLC signal transmission safe and secure. However, these ML models are hugely dependent on the data provided at the training stage for accurate identification of the legitimate operation or any anomaly behavior of the wireless networks. As the wireless traffic is growing exponentially, the volume of data generated from this process is rapidly increasing. In future 6G, managing this high volume of data and providing the correct set of data for training the ML models will become crucial. As mentioned above, various mission-critical HRLLC applications will depend on ML-enabled techniques to automate operations and achieve secure data transmission. Therefore, in case of anomaly, we should have the capability to delete the suspicious data points to retrain the ML models for obtaining secure prediction results. In this regard, machine unlearning is a potential new technique introduced recently to mitigate the effect of malicious data points on the performance of ML-enabled URLLC/HRLLC operations in upcoming 6G wireless networks.

Machine unlearning is an innovative process of removing sensitive information or data points from a trained ML model \cite{ul1}. This procedure is typically carried out to address privacy issues, adhere to data protection laws, or eliminate model biases. Machine unlearning is required in a number of circumstances such as, to change or eliminate data points that were accidentally included to the training set in cases where sensitive or private information was unintentionally included. Unlearning enables the model to adapt to new data patterns and avoid detrimental consequences from out-of-date knowledge in dynamic contexts where data distributions change over time \cite{ul2}. Rather than completely retraining a model, unlearning can be used to incrementally update a model by adding or removing individual data points. To safeguard privacy during unlearning and maintain model correctness, methods like differential privacy, secure multi-party computation, or federated learning are frequently utilized.

Wireless communication systems can benefit from machine unlearning in a number of ways, particularly with regard to data privacy, security, and adaptability. In wireless communication networks, confidential data such as user location information and user identity may be gathered for a number of purposes. In such a scenario, privacy issues may arise if sensitive information is unintentionally included in the training process. In order to protect user privacy and adhere to data protection rules, machine unlearning enables the removal of such sensitive data from the learned models. The wireless communication environment is dynamic in nature due to the presence of noise and interference, network congestion, and fluctuating channel conditions. Therefore, machine unlearning can be used to update ML models and delete old or unnecessary data that hamper the overall performance and helps the model to adapt to the dynamic environment. Additionally,  machine unlearning can assist in addressing bias in ML models that might unfairly disadvantage some user groups. The model can be changed to offer fair and equal services to all users by deleting biased or discriminatory data points \cite{ul3}.

\begin{figure}[t]
	\centerline {\includegraphics[width=1\linewidth, height=6.1 cm]{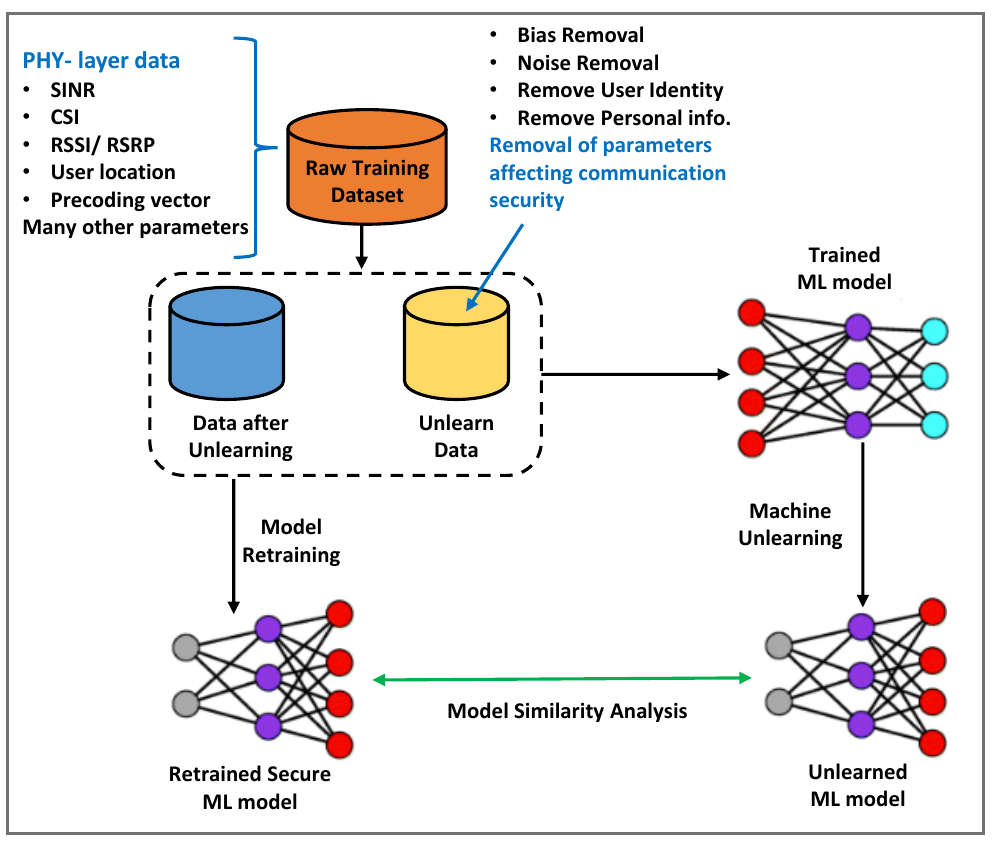}}
	\caption{Illustration of Machine Unlearning Process for PLS}
	\label{fig11}
	\vspace{-4 mm}
\end{figure}

Moreover, wireless communication systems may be subject to attacks from eavesdroppers that try to harm the system's functionality and intercept confidential information. The resistance to attacks can be improved by removing data points that have been altered or disrupted by attackers via unlearning. Therefore, machine unlearning can be utilized successfully to remove data points that have been manipulated by active eavesdroppers and enhance the resilience of the ML model to eavesdropping attacks. At the end of the day, machine unlearning can be a useful tool for improving the fairness, security, adaptability, and privacy of URLLC/HRLLC systems, opening the door for more ethical and responsible usage of these technologies.

\begin{figure*}[t]
	\centering
	\includegraphics[width=0.95\linewidth, height=11.5 cm]{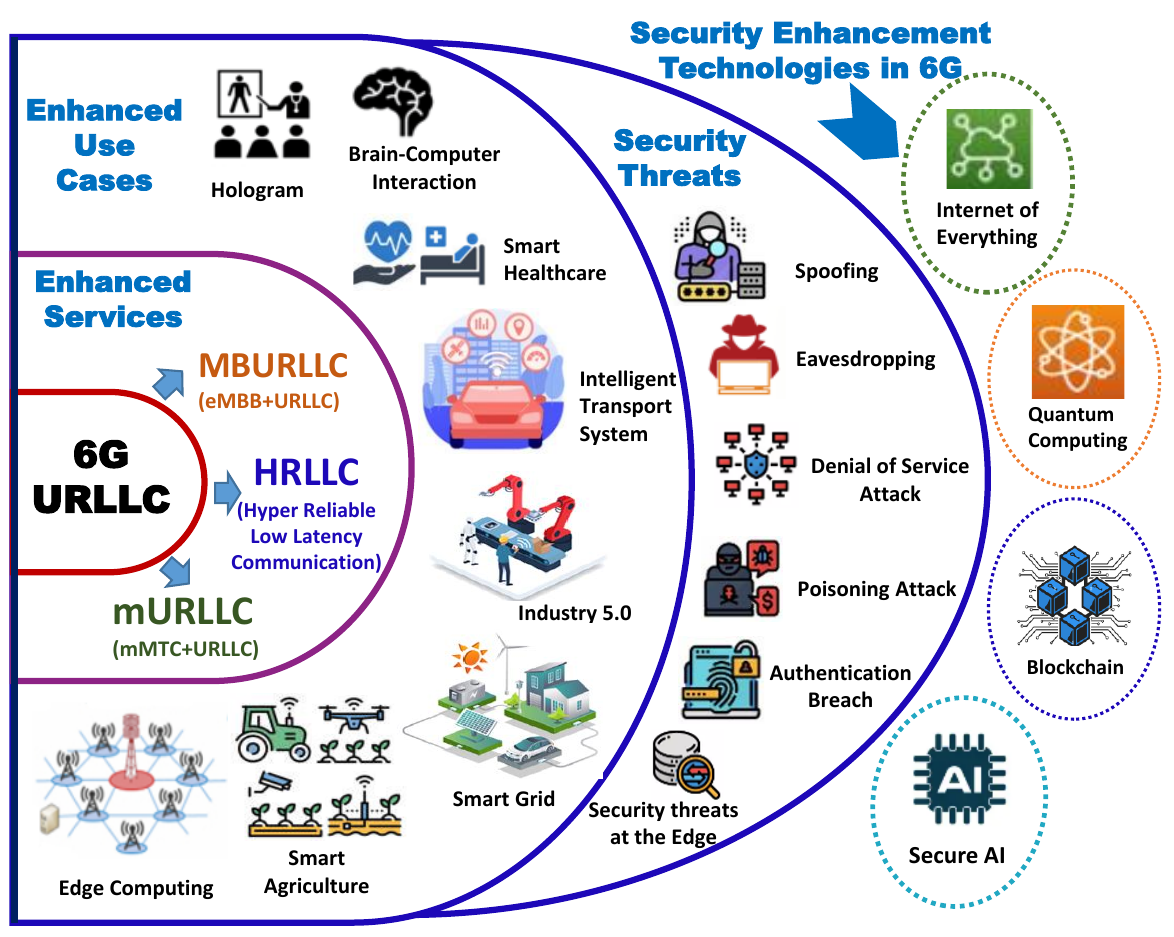}
	\caption{6G HRLLC service, Use cases, Security threats, and Enabling Technologies}
	\label{fig10}
	\vspace{-4mm}
\end{figure*}

\subsection{Blockchain enabled Security in 6G}

Conventional ML system design adopts centralized architecture to provide global optimal solutions for managing multiple tasks simultaneously. However, the centralized collection of data, computational complexity, and large processing time make these centralized ML models unsuitable for URLLC operations \cite{r151}. Moreover, the centralized approach often leads to a lack of trust om ML models and security vulnerability in URLLC data transmission. Basically, distributed architecture is suitable for URLLC users in a large-scale network or at the edge computing scenario. The edge URLLC users or distributed URLLC users train the local ML models with their datasets independently and only share the model updates with the central model \cite{pls63}. This ensures the confidentiality of the user data. As we have discussed earlier, the federated learning models offer the best management of resources and task offloading for URLLC users, which also potentially preserves data privacy using a decentralized, distributed architecture. However, the major concern is to find a proper decentralized security solution that can address the security and privacy issues of URLLC service. One such potential solution is the use of blockchain to design decentralized security-preserving ML models. 

 The basic operation of a blockchain technique involves hashed data stored in an immutable and read-only data format. The blocks are appended to the ledger by using the hash identifier of the previous block, making it a more trustworthy and secure data-sharing method \cite{pls4}. Moreover, using blockchain can provide an extra degree of security to URLLC data transmission with the help of timestamped and tamper-proof shared parameters \cite{r143}, \cite{pls3}. Therefore, the futuristic mission-critical 6G HRLLC applications will hugely benefit from the customized blockchain techniques in facilitating secure data transactions.

However, the major critical issue for URLLC/HRLLC service is the low latency criteria. On one side, the decentralized architecture of the blockchain technique makes data secure for URLLC/HRLLC applications; on the other side, the transactions and smart contracts used to generate metadata in the blockchain may disclose user personal information. Although blockchain provides immutable secrecy, the data encryption process used in blockchain may increase the overall transmission latency for URLLC/HRLLC. Therefore, designing efficient secure data transmission schemes using blockchain and maintaining the QoS constraint of URLLC/HRLLC will be an open issue for the researchers.

\subsection{Green physical layer security}

Power-efficient secrecy transmission strategies need to be developed for URLLC/HRLLC to support massive communication scenarios integrating a large number of low-power IoT devices in 6G wireless networks. It is noteworthy that, strict secrecy constraints lead to a decrease in the secrecy rate of URLLC/HRLLC. Hence, high power allocation is one of the key ways to boost the SNR of legitimate users and improve the achievable secrecy rate \cite{pls10}. For improving PLS against eavesdropping, the cooperative relaying technique has been gaining attention in recent years. However, the involvement of low-power IoT devices and energy-limited relay nodes in cooperative communication like UAVs, and IRS introduces challenges in designing energy-efficient PLS schemes.

In 6G, it will become significant to consider the energy limitations, security constraints, and QoS requirements of HRLLC for effective PLS design for a large number of low-power end users. Therefore, the development of green PLS techniques is highly desirable for ensuring energy-efficient sustainable secure communication in future 6G HRLLC applications where massive low-power devices are involved. Moreover, the evaluation of power consumption at the device level, circuit level, and network level will be imperative for achieving green PLS while enabling HRLLC service in 6G.

\section{Open Challenges and Future Research Directions}

Secure communication of URLLC signals is the fundamental requirement of 5G and future wireless networks. There are various factors on which the PLS of URLLC is dependent, and it is widely discussed in the literature. The challenges enforced by URLLC system design parameters like strict reliability, low latency, non-zero error probability, and limited blocklength on the PLS design amplify the complexity. In addition, the trade-off between security and URLLC QoS limits the possibilities of designing efficient security countermeasure techniques. After reviewing a significant amount of surveys and research contributions on URLLC and PLS techniques utilized in wireless communication, we feel the need for widely accepted security standards and evaluation metrics, which were missing in most of the cases. Moreover, there is a lack of in-depth study, analysis, and discussion on the possible enhancements of the PLS schemes for future 6G URLLC applications.

Apart from the above observations, we provide a detailed summary and critical analysis of the aforementioned sections containing PLS methods adopted for URLLC use cases, enabling techniques, and ML-enhanced approaches. Not limiting this discussion, we also give a more futuristic perspective to this survey by identifying the key challenges in providing secure URLLC for 6G and some exciting new future technologies that can result in really groundbreaking solutions and research opportunities. Moving forward with our discussion, we aim to expose various open challenges in ensuring PLS for URLLC/HRLLC service and list out potential future research directions for the researcher community.

\subsection{Open Research Challenges}
Ensuring the physical layer security of latency-critical information in wireless communication is a critical issue. As large-scale wireless networks become more prevalent in various 5G applications, the challenges of providing secure communication continue to emerge. These challenges require innovative security enhancement solutions to address potential threats and vulnerabilities. In this section, we identify some open research challenges in providing the PLS for latency-sensitive URLLC service as listed below.
   
\subsubsection{Stringent QoS and URLLC design parameters }
Generally, URLLC applications require data transmission under stringent low latency requirements. Moreover, finite blocklength codes for URLLC services introduce challenges for designing effective security solutions. Due to FBL codes decoding error at the receiver is unavoidable. On the other hand, it increases the chances of interception of confidential information. Therefore, the design of lightweight PLS solutions under the extreme QoS requirements of URLLC is highly desired.

\subsubsection{Dependency on CSI}
The CSI-based PLS schemes are widely adopted for providing security to URLLC signal transmission. However, imperfect or partial CSI knowledge will degrade the performance of PLS schemes as they result in sub-optimal security solutions for URLLC. Moreover, considering statistical CSI for developing PLS schemes for URLLC may not provide efficient PLS due to harsh wireless environmental conditions. It shows that the secure transmission of URLLC signal will be hugely affected by the lack of availability of accurate characterization of the wireless environment and standardized channel models. Therefore, accurate CSI estimation and PLS scheme design without depending on CSI is a bottleneck for URLLC.

\subsubsection{Edge Computing}
The reliable communication for edge devices introduces security vulnerability due to large transmitter-receiver separation distance and weak channel conditions. In such cases, localized processing and caching of the information is done to minimize processing and transmission delays.  However, the information transfer between the central server and the edge devices is affected by channel fading and information leakage. Moreover, short packet communication may get hampered by DoS attacks at the central server or unwanted intrusion. Hence, lightweight PLS enhancement techniques with faster authentication and integrity checks need to be developed for providing secure communication to the URLLC/HRLLC users at the cell edge.

\subsubsection{Growing network traffic and service demand}
The growing data traffic and service demand in large-scale IoT networks bring challenges in optimizing the PLS technique for URLLC/HRLLC signal transmission. Additionally, the dynamic nature of wireless channels introduces roadblocks to the design of efficient PLS schemes. Moreover, maintaining the confidentiality of these large volumes of data requires low complexity and flexible security solutions. Therefore, scalable and flexible security solutions have to be developed to address the aforementioned challenges.

\subsubsection{Security of network slicing}
The complexity of radio resource management increases in various mission-critical applications due to the coexistence of heterogeneous service types \cite{pls100}. The allocation of virtually sliced wireless network resources can provide efficient radio resource management for the coexisting 5G services with respect to individual QoS requirements \cite{pls94}. However, these network slices have different security constraints based on heterogeneous service requirements. Therefore, developing service-aware adaptive security levels is another challenging task for next-generation security systems in 6G.  

\subsubsection{User mobility}
There are very limited studies present that consider the mobility of URLLC UEs while designing PLS solutions. Mobility is an important factor to be considered for URLLC use cases like smart transportation, automated guided vehicles in industries, and UAV-aided communication. Therefore, it becomes essential to study the impact of mobility on the secrecy constraints and the physical channel parameters at the legitimate receiver and the eavesdropper. Therefore, it is an open issue for the researchers to propose efficient PLS schemes for URLLC under user mobility.

\subsubsection{Limitations of ML}
ML-based techniques have been extremely useful for the identification of anomalous behaviors and malicious attacks in wireless networks. However, the use of adversarial ML techniques by attackers introduces new challenges for secure communication. Therefore, it becomes essential to develop ML models that are resilient to these adversarial attacks and can effectively detect and mitigate security threats in URLLC/HRLLC applications.

\subsubsection{Research gaps in emerging technologies}
It is crucial to differentiate various URLLC use cases based on heterogeneous QoS requirements and enabling techniques by analyzing the security constraints. This will allow the design of more robust practical boundaries for PLS solutions considering real-time wireless environment scenarios of mission-critical URLLC use cases. New technologies like IRS and UAV-assisted communication provide reliable and energy-efficient signal transmission for URLLC. For that, they often require the position information of URLLC UEs and eavesdropper location. The knowledge of UE position can increase vulnerability and compromise the optimality of the proposed PLS scheme. The research contributions mentioned in Section IV provide some attractive solutions for ensuring the secure transmission of URLLC signals with the help of relays like UAVs and IRS. However, we feel the requirement for further research in this domain to design PLS schemes without revealing the user location information.

\subsection{Future Research Directions}

Ensuring wireless communication security in future URLLC/HRLLC applications needs to address the evolving landscape of security threats and the exponential rise in wireless network traffic by facilitating technological advancements for secure data transmission. In this regard, the physical layer holds untapped potential for security enhancements with the help of innovative physical layer techniques like optimal resource allocation, adaptive beamforming, advanced waveform modulation, and channel encoding to preserve the confidentiality of the short packet URLLC/HRLLC signal transmissions from eavesdropping and jamming attacks. Therefore, future research endeavors need to emphasize the development of lightweight, dynamic, and energy-efficient PLS protocols that can address the inherent challenges and performance trade-offs for URLLC/HRLLC service. 

In summary, to provide a holistic research outlook, we review the recent advancements in developing efficient PLS schemes for URLLC/HRLLC and enlist some of the potential future research directions for ensuring the PLS of URLLC/HRLLC service in future 6G wireless networks as follows.

\begin{itemize}
  
  \item
Accurate CSI estimation under URLLC QoS constraints is one of the challenging tasks for ensuring PLS. Thus, efforts need to be oriented in the direction of achieving more accurate channel models with correct CSI estimation. Moreover, reducing the dependency on the CSI while developing efficient PLS schemes for URLLC will be an interesting future research direction.

\item
Relay-assisted communication using UAV and IRS improves the reliability and security of URLLC. However, the extra power consumption by the relay nodes degrades the overall energy efficiency of the system. Therefore, designing PLS schemes with energy harvesting and developing green PLS schemes will be one of the important considerations for HRLLC applications in the future. 

\item
An interesting future research direction can be the exploration of ML techniques for designing intelligent PLS techniques for enabling secure URLLC. There is not much work available yet on ML-aided PLS design. Therefore, integration of both ML and PLS techniques exploiting the physical layer features for secure HRLLC will be a key consideration for 6G wireless networks.

\item
Recently, federated learning has emerged as the potential ML model for preserving distributed training data locally and the privacy of edge users' information. However, malicious edge participants can deliberately interfere and intercept the information in FL. Therefore, the blockchain-based FL framework addresses the issue by preventing unreliable participants from intercepting and leaking private information.

\item 
The secure communication requirement for large-scale wireless networks in delay-sensitive URLLC applications will alter with the introduction of new technologies like quantum computing. The development and standardization of security enhancement techniques which can withstand quantum attacks will be the main topic of future research. Exploring the possible integration of quantum-safe solutions to wireless protocols for secure communication in 6G will be crucial to ensure long-term security. Moreover, PLS enhancement by combining blockchain and quantum communication techniques can be explored in the future to support massive connectivity and security for URLLC applications in 5G and HRLLC in 6G simultaneously.

\end{itemize}

Exploring these future research avenues will not only improve the performance of security enhancement schemes but also fill the current gaps in technology to facilitate more reliable, and secure URLLC/HRLLC in the 6G era.

\section{Conclusion}

This survey paper addressed the security aspect of URLLC/HRLLC signal transmission from a physical layer perspective, which is one of the essential requirements to enable secure data transmission in mission-critical applications. We first provided a comprehensive overview of PLS for finite blocklength URLLC, in which the impact of various system parameters like CSI value, signal blocklength, and pilot signal length along with the security threat landscape at the physical layer was succinctly introduced. Then we discussed the PLS performance evaluation metrics and the state-of-the-art PLS techniques utilized to ensure secure URLLC in typical 5G mission-critical applications. In addition, the survey presented a detailed study of PLS techniques adopted in emerging URLLC enabling 5G technologies (i.e., MIMO, mMIMO, NOMA, relay assisted, and cooperative communication) along with a review of related research contributions for providing in-depth understanding. Apart from this, we discussed the role and benefits of ML-enabled security solutions for URLLC from both physical layer and cross-layer perspective to provide a holistic approach for future applications. Most importantly, the survey shaded light on the future security aspects of the 6G envisioned new service category HRLLC while discussing the role of various key emerging technologies like quantum communication, machine unlearning, blockchain, and green PLS techniques in providing robust PLS design for URLLC and 6G HRLLC applications. We also identified several research challenges and open issues in implementing PLS for URLLC that need to be addressed carefully while developing efficient consensus PLS protocols to manage service trade-offs efficiently. 

In summary, providing secure communication for extremely delay-sensitive HRLLC applications in 6G needs the development of efficient and adaptive PLS schemes by combining innovative security solutions along with real-time threat detection. With the help of these research initiatives, it will be possible to fulfill the transformational potential of HRLLC applications without sacrificing security and reliability. Therefore, in this survey, we highlight some of the new emerging research directions for facilitating secure URLLC/HRLLC service in 6G wireless networks, which will provide new and exciting future research avenues for wireless communication researchers. We hope this survey will provide the required fundamentals for further research on the security of mission-critical wireless communication.

\bibliographystyle{IEEEtran}
\bibliography{main.bib}
\vskip -2\baselineskip plus -1fil
\begin{IEEEbiography}[{\includegraphics[width=1in,height=1.25in,clip,keepaspectratio]{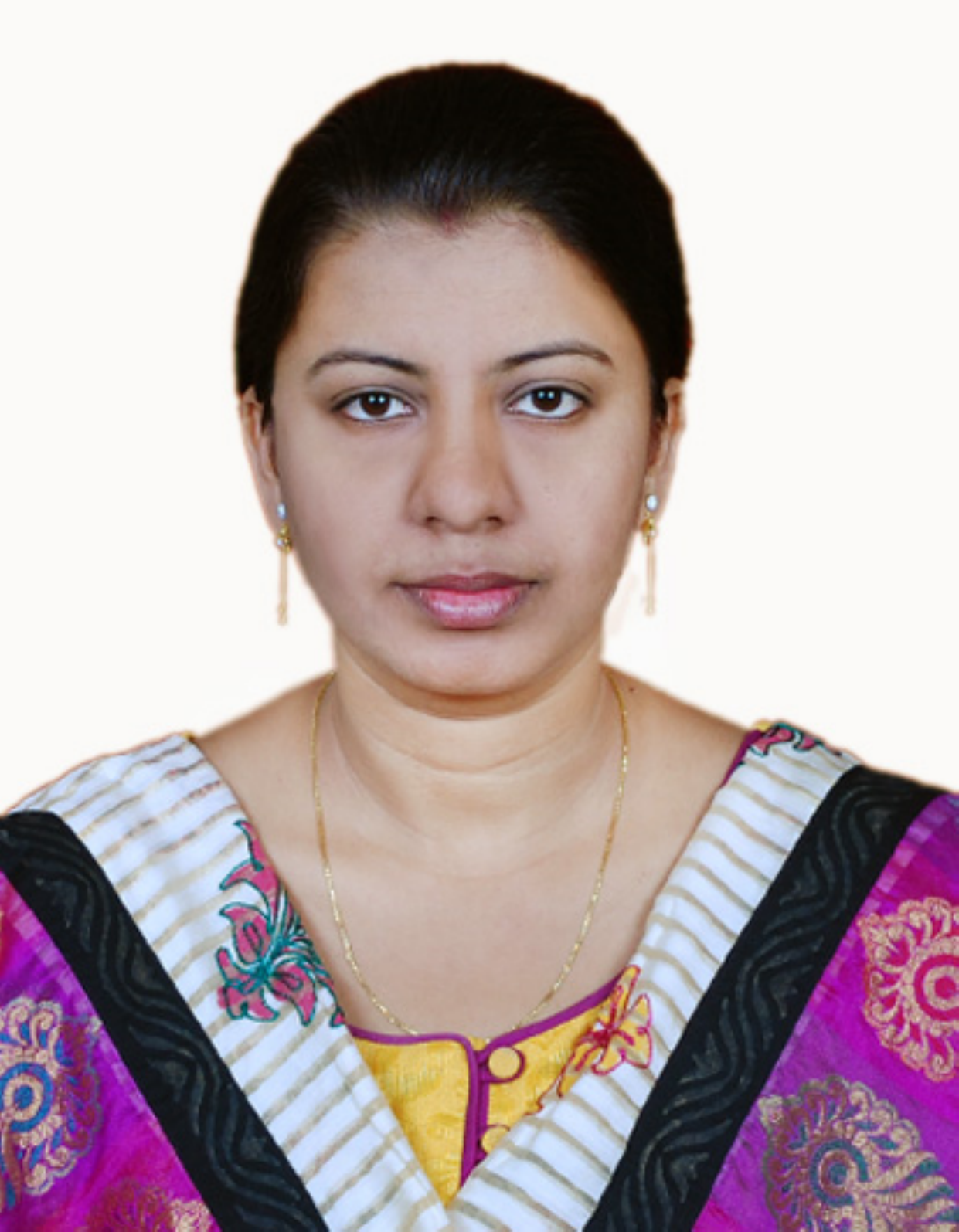}}]{Annapurna Pradhan} received her BTech and MTech degree in Electronics and Telecommunication Engineering from Biju Patnaik University of Technology, India in 2008 and 2012, respectively. She is currently pursuing her Ph.D. degree in Electrical Engineering at the National Institute of Technology, Rourkela, India. She was the recipient of the "Young Investigator" award and the "Best Paper" award at ICMLCI-2019 for her work. Her research interests include Ultra-reliable low-latency communication, physical layer security, NOMA, machine learning, and quantum communication.
\end{IEEEbiography}

\vskip -2\baselineskip plus -1fil

\begin{IEEEbiography}[{\includegraphics[width=1in,height=1.25in,clip,keepaspectratio]{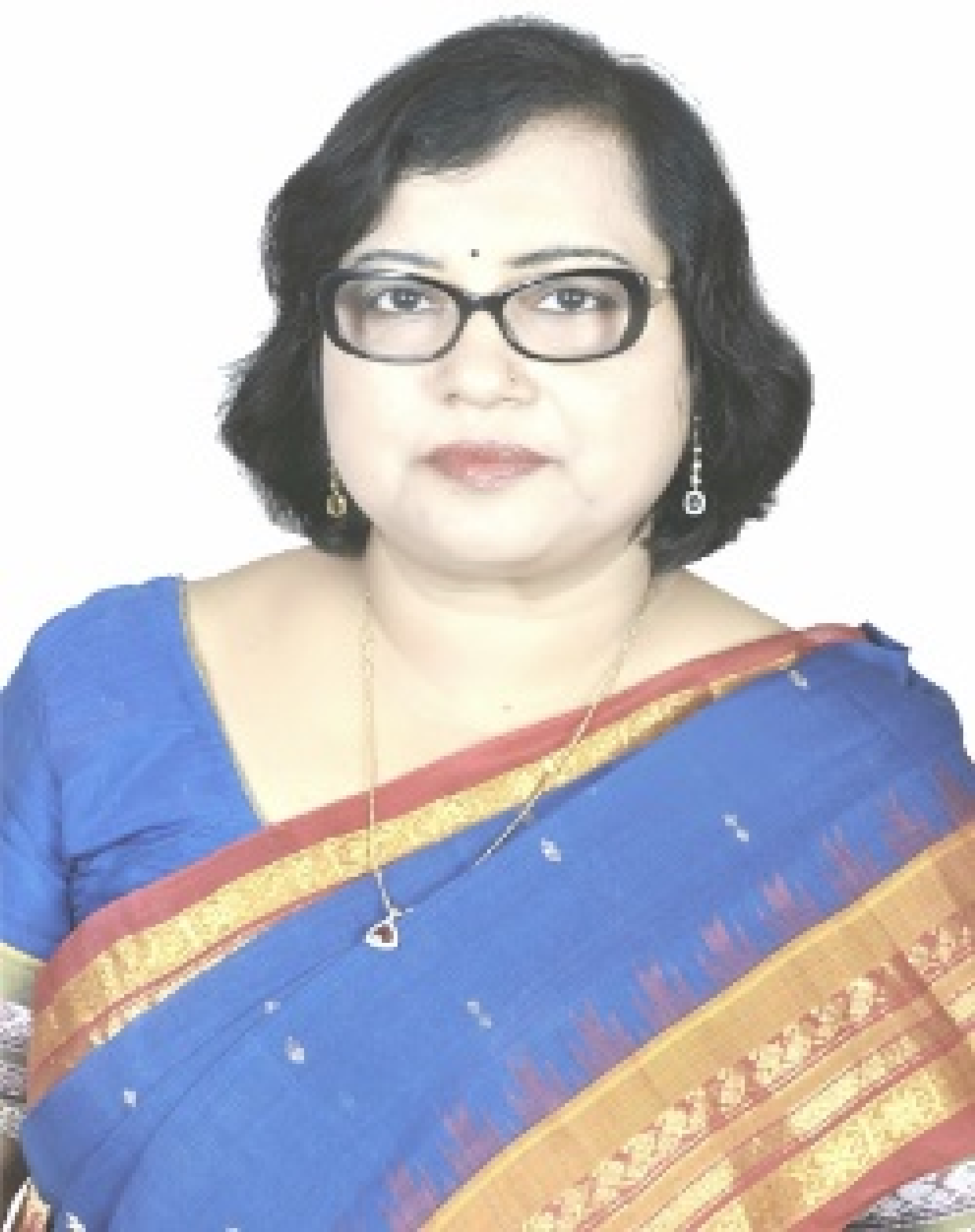}}]{Susmita Das}  (M'10, SM'16) is currently serving as a Professor in the Department of Electrical Engineering, National Institute of Technology, Rourkela, India. She received her BSc degree from the College of Engineering and Technology, Odisha, India in 1989. She earned her MTech. and Ph.D. in Electrical Engineering from the National Institute of Technology, Rourkela, India in 1995 and 2004, respectively. She has authored and co-authored more than 120 technical research papers in leading journals and conferences. Prof. Das holds a distinguished academic background and was the recipient of several Best Paper awards for her work. Her research interests include adaptive beamforming, URLLC, cooperative relaying, D2D communication, IRS, UAV, adaptive signal processing, Cognitive Radio, and Machine Learning. She is a Senior Member of IEEE, a Member of IETE, a Life member of ISTE, and a life member of the Institute of Engineers, India.
\end{IEEEbiography}	

\vskip -2\baselineskip plus -1fil

 \begin{IEEEbiography}[{\includegraphics[width=1in,height=1.25in,clip,keepaspectratio]{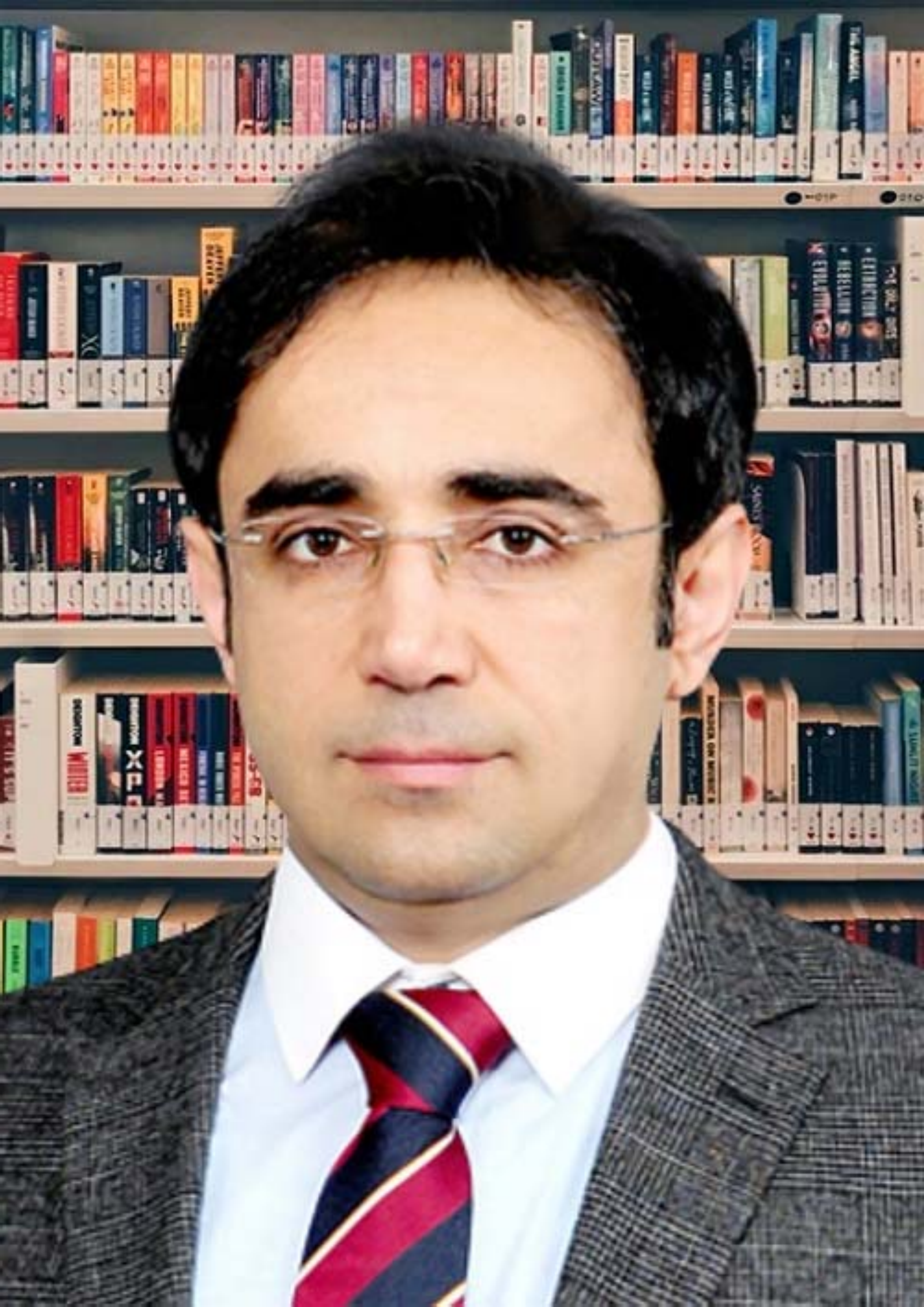}}]{Md. Jalil Piran (S'10, M'16, SM'21)} holds a distinguished academic background and currently serves as an Associate Professor at the Department of Computer Science and Engineering in Sejong University, Seoul, South Korea. He earned his Ph.D. degree in Electronics and Information Engineering from Kyung Hee University, South Korea, in 2016, and subsequently worked as a Post-Doctoral Fellow at the Networking Laboratory of the same institution.

Prof. Piran has made significant contributions to the field of Artificial Intelligence and Data Science through his extensive research publications in esteemed international journals and conferences. His areas of expertise encompass Machine Learning, Data Science, Big Data, Internet of Things (IoT), and Cyber Security. In addition to his research endeavors, Prof. Jalil Piran actively engages with scholarly journals as an Editor, including the "IEEE Transactions on Intelligent Transportation Systems," "Elsevier Journal of Engineering Applications of Artificial Intelligence," "Elsevier Journal of Physical Communication," and "Elsevier Journal of Computer Communication." He also serves as Secretary of the IEEE Consumer Technology Society on Machine Learning, Deep Learning, and AI. Furthermore, he assumes the role of Track Chair for Machine Learning, Deep Learning, and AI in the CE (MDA) Track for the upcoming 2024 IEEE International Conference on Consumer Electronics (ICCE). In 2022, he chaired the "5G and Beyond Communications" Session at the prestigious IEEE International Conference on Communications (ICC). 
Prof. Piran is esteemed as a Senior Member of IEEE and represents South Korea as an Active Delegate to the Moving Picture Experts Group (MPEG). His outstanding research contributions have been recognized internationally, as evidenced by the prestigious "Scientist Medal of the Year 2017" awarded by IAAM in Stockholm, Sweden. Moreover, he received accolades from the Iranian Ministry of Science, Technology, and Research as an "Outstanding Emerging Researcher" in 2017. His exceptional Ph.D. dissertation was honored as the "Dissertation of the Year 2016" by the Iranian Academic Center for Education, Culture, and Research in the Engineering Group.
\end{IEEEbiography}

\vskip -2\baselineskip plus -1fil

\begin{IEEEbiography}[{\includegraphics[width=1in,height=1.25in,clip,keepaspectratio]{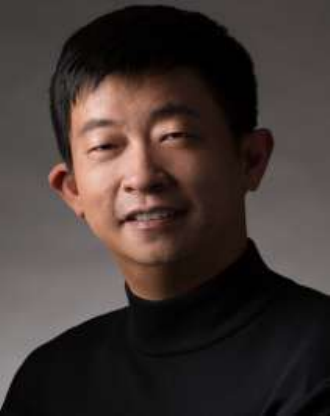}}] {Zhu Han} (Fellow, IEEE) received the B.S. degree in electronic engineering from Tsinghua University, in 1997, and the M.S. and Ph.D. degrees in electrical and computer engineering from the University of Maryland, College Park, in 1999 and 2003, respectively. From 2000 to 2002, he was an R$\&$D Engineer at JDSU, Germantown, Maryland. From 2003 to 2006, he was a Research Associate at the University of Maryland. From 2006 to 2008, he was an assistant professor at Boise State University, Idaho. Currently, he is a John and Rebecca Moores Professor in the Electrical and Computer Engineering Department as well as in the Computer Science Department at the University of Houston, Texas. Dr. Han’s main research targets on the novel game-theory related concepts critical to enabling efficient and distributive use of wireless networks with limited resources. His research interests include wireless resource allocation and management, wireless communications and networking, quantum computing, data science, smart grid, security, and privacy. Dr. Han received an NSF Career Award in 2010, the Fred W. Ellersick Prize of the IEEE Communication Society
in 2011, the EURASIP Best Paper Award for the Journal on Advances in Signal Processing in 2015, IEEE Leonard G. Abraham Prize in the field of Communications Systems (best paper award in IEEE JSAC) in 2016, and several best paper awards in IEEE conferences. Dr. Han was an IEEE
Communications Society Distinguished Lecturer from 2015-2018, AAAS fellow since 2019, and ACM distinguished Member since 2019. Dr. Han is a 1$\%$ highly cited researcher since 2017 according to Web of Science. Dr. Han is also the winner of the 2021 IEEE Kiyo Tomiyasu Award for outstanding early to mid-career contributions to technologies holding the promise of innovative applications, with the following citation: “for contributions to game theory
and distributed management of autonomous communication networks.”
\end{IEEEbiography}

\end{document}